\documentstyle[12pt,a4]{article}
\begin{document}
\baselineskip 21pt
\begin{center}
{\large {\bf CASCADE $\gamma$-DECAY OF THE $^{193}Os$ COMPOUND NUCLEUS
AND SOME ASPECTS OF DYNAMICS OF CHANGE IN NUCLEAR PROPERTIES BELOW $B_n$}}
\end{center}
\begin{center}{\bf VALERY.A.BONDARENKO$^a$, JAROSLAV.HONZ\'ATKO$^b$, VALERY. A. KHITROV$^c$, ANATOLY. M. SUKHOVOJ$^c$, IVO.TOMANDL$^b$}\\
{\sl $^a$Nuclear Research Center, LV 2169 Salaspils, Latvia}\\
{\sl $^b$ Nuclear Physics Institute, CZ-25068 \v{R}e\v{z} near Prague, Czech Republic}\\
{\sl $^c$Frank Laboratory of Neutron Physics, Joint Institute for Nuclear
Research \\ 141980 Dubna, Russia}
\end{center}

\begin{center}
UDC 539.172.4
\end{center}
\begin{center}
\end{center}

Two-step cascades from the $^{192}Os(n_{th},\gamma)^{193}Os$ reaction were studied in
$\gamma - \gamma$ coincidence measurement. The decay scheme of $^{193}Os$ was established up to the
excitation energy $\sim 3$ MeV. The excitation spectrum of intermediate levels of most intense cascades was found to be practically harmonic.

\section{Introduction}
Nuclear properties in the excitation interval up to the neutron binding
energy $B_n$ undergoes radical change: the simplest low-lying levels
transform into the Bohr's compound states. The only possibility to study
this process in details is provided by the experimental investigation of the
two-step $\gamma$-cascades proceeding between the neutron resonance and
group of low-lying levels. Experimental data on the density and probability
of population of the states observed in this process are compared with the
theoretical notions. The comparison allows one to reveal the main
peculiarities of change in properties of nuclear matter as excitation energy
increases. The main details of the analysis of this experiment are described,
for example, in [1-3].

\section{Experiment}

Two-step $\gamma$-cascades following thermal neutron  capture in
$^{190}Os$ and $^{192}Os$ were studied by $\gamma$-$\gamma$ coincidence
measurements undertaken at the $LWR-15$ reactor in \v{R}e\v{z}. The
measurements were performed using the spectrometer [4] consisting of two
$HPGe$-detectors with the efficiency 20\% and 30\%. The target consisting of
1200 mg of $^{192}Os$ and 176 mg of $^{190}Os$ was used. As the thermal
neutron capture cross section [5] equals 13.1 b for $^{190}Os$ and 3.12 b for
$^{192}Os$, then this target provided 38\% of captures in $^{190}Os$
and 62\% in $^{192}Os$.

Unlike other known methods for the study of the process of thermal neutron
capture, the sum coincidence method allows one to obtain reliable enough
information not only for a monoisotopic target but also for the case of few
isotopes with comparable probabilities of neutron capture in them. In
the last case, the quality of the experimental data is somewhat worse due to:

(a)~increase in the Compton background under the full energy peak in
the sum coincidence spectrum caused by a higher-energy cascade belonging
to another isotope;

(b)~possible overlapping of peaks in the sum coincidence spectrum.\\
However, a sufficiently high efficiency of detectors and fine energy
resolution ($FWHM\simeq 5$ keV for peaks at $E_c=5-6$ MeV in the sum
coincidence spectrum) allowed us to obtain the results of acceptable quality
in this case, as well.

The main part of the sum coincidence spectrum measured in the experiment
is shown in Fig.~1. A relatively large background under the peaks in the
region of the neutron binding energy, $B_n$, is caused by a parasitic
neutron capture in $\simeq 3$ mg of $Cl$ contained in the target and
surrounding constructions. This component of the background and Compton
background at the lower cascade energy $E_c=E_1+E_2$ determine amplitude
and shape of the ``noise" line in the intensity distributions of cascades
with $E_c=const$. In all there were obtained and analysed 11 such spectra
of cascades in $^{193}Os$. Each two-step cascade in such
spectrum is presented by a pair of peaks with equal areas and widths [6].
The probability of observing a low-intensity cascade is
determined only by the amplitude of the ``noise" line. The registration
threshold, $L_c$, for individual cascades was determined from an analysis of
spectra  $E_c=const$ corresponding to background intervals
in the sum coincidence spectrum. It was established that $L_c$
linearly increases from 1.5 to 6.0 events per $10^4$ decays as the cascade
energy changes from 5.6 to 4.5 MeV, respectively.

The data obtained allow one clearly demonstrate spectroscopic possibilities
of the method. For this aim all 11 intensity distributions of cascades
observed were summed. The part of this summed spectrum for the interval of
the quantum energy $E_{\gamma}$
from 3.0 to 4.6 MeV is presented in Fig.~2 in function of the energy
$E_{ex}=B_n-E_{\gamma}$ ($E_{ex}$ equals excitation energy of the cascade
intermediate level when $E_{\gamma}$ is the energy of the cascade primary
transition $E_1$). Summation leads to accumulation of the primary transition
intensities in common peak and intensities of the secondary transitions --
in different peaks. Due to low intensity cascades with the secondary
transition energy $E_2>$ MeV are not observed in a given interval of the
excitation energy as the peaks, they form continuous distribution. Therefore,
every peak in Fig.~2 with the high probability corresponds to one of the
levels of $^{193}Os$. So, one can see from Fig.~2 that the experimental data
allow confident determination of the level energies in $^{193}Os$ up to
$0.5B_n$ as minimum even without the use of the most modern spectrometers.

\section{Spectroscopic information}
The method to construct a decay scheme using obvious thesis about the
constancy of the energy $E_1=B_n-E_m$ of the primary transition in the
cascades with the different total energy $E_1+E_2=const$ was described
for the first time in [7]. The method uses multi-dimensional distribution
in the framework of the maximum likelihood method in order to select
probable $\gamma$-transitions with equal energy in different spectra.
The algorithm gives reliable results [8] even at the mean error in
determination of $E_1$ up to $\simeq 1.5$ keV and number of cascades resolved
in spectrum as the pairs of peaks of about $10^3$. Corresponding results for
$^{193}Os$ are given in Table~1.

Analysis of the experimental data requires transformation of the peak
areas of the resolved cascades into absolute values (in \% per decay).
However, the direct solution of this problem using, for example, the areas
of peaks in the sum coincident spectrum is impossible because of the
uncontrollable conditions of the experiment. First of all,
this is due to difficulties of determining the number of captures in the
target and the absolute efficiency of registration of the cascade in the
geometry of the experiment. This problem can be solved by the normalization
of relative intensities to the absolute values $A_{\gamma\gamma}$ calculated
for most intense cascades by the relation
\begin{equation}
A_{\gamma\gamma}=i_1\times B_r,
\end{equation}
where the absolute intensities $i_1$ of primary transitions are taken
from other works, and the branching ratios $B_r$ are determined in a
standard way from the codes of coincidences accumulated in this experiment.
$A_{\gamma\gamma}$ is the ratio between the intensity of a given cascade
and total intensity of all cascades with $E_1+E_2=const$. The use of a
maximum large ensemble of reference cascades in the normalization allows
one to minimize both statistical and systematical errors of the procedure
and practically reduce them to existing  errors of $i_1$.

Unfortunately, there are no reliable data on the absolute intensities of
primary transitions for osmium isotopes under consideration. Therefore,
we were forced to use the data [9] on relative intensities of
$\gamma$-transitions following thermal neutron capture in $^{190}Os$. For
their normalization, we measured the spectrum of $\gamma$-rays after thermal
neutron capture in the target of natural $Os$ and determined ratios between
the peak areas corresponding to $\gamma$-transitions with energies 7234,
7793($^{190}Os$), 7835($^{188}Os$), and 5147 keV ($^{191}Os$). The absolute
intensity of 5147 keV transition belonging to $^{191}Os$ was determined to be
equal $I_1=14.4(14)\%$ per decay using absolute intensities [10,11] of three
other transitions and data [5,12] on isotopic abundance  and thermal
neutron capture cross sections. This allowed us to reduce relative
intensities of primary transitions in $^{191}Os$ [9] into the absolute values
and determine [13] intensities of the two-step cascades in this nucleus.
The coefficient of normalization of relative intensities of the primary
transitions in $^{193}Os$ to absolute values was determined using the ratio
of areas of the peaks corresponding to cascades to the ground and three
low-lying levels of $^{191,193}Os$ as well as the averaged over corresponding
spectra efficiency of registration of the cascades. It was determined to
be equal to 0.01077 for the relative intensities of the primary transitions
from [14].

The total absolute intensities $I_{\gamma\gamma}=\sum{i_{\gamma\gamma}}$ of
cascades with a fixed sum energy (including those unresolved experimentally)
are given in Table~2. The data correspond to the energy detection threshold
set at 520 keV which was used to reject annihilation quanta. Nevertheless,
the data are suitable for testing the validity
of level density and radiative strength function models in the excitation
energy range almost up to $B_n$ as it is shown in Table~2.

\subsection{Background cascades}
Every spectrum -- intensity distribution of cascades with a given sum
energy contains the following components:
(i)~desired cascade  both in form of pairs of resolved peaks and their
superposition -- continuous distribution of low amplitude (cascade energy
is completely deposited in detectors);
(ii)~``noise" line resulted from registration of the part of the energy
of quanta of cascades with higher energy.

On the average, the contribution of the latter in the spectrum practically
equals zero but in some local sections of the spectrum the distortions can
be rather considerable. It should be noted that the main distortion
is due to the cases of partial absorption of the energy of one cascade
transition in detector and complete absorption of the energy of another
transition. Subtraction of this Compton background results
in appearance of characteristic symmetrical structures of variable sign
in spectrum of cascades with less sum energy. As it was shown in [8]
these structures manifest themselves when the full energy peaks of
high-energy cascades contain more than 1000 events. The use of the
numerical algorithm for improvement of energy resolution [6] strengthens
this effect. The shape of such structure is determined by the intensity
of corresponding individual cascade with higher energy, choice of concrete
windows ``effect+background" and ``background" (Fig.~1) and is due to
inevitable discrepancy in positions of these intervals. Of course, similar
effect exists and in the standard analysis of $\gamma -\gamma$ coincidences
but there it is well ``masked".

Precise enough and complete correction of corresponding distortions in the
spectra, from which the data of Fig.~2 were obtained, can be calculated.
This requires the data on the probability of simultaneous registration of
quanta $E_1$ and $E_2$ in the full energy peaks for all possible most
intense cascades with the higher energy (included cascades in other
isotopes and elements situated in neutron beam). This procedure is
absolutely necessary if $HPGe$ detectors with the efficiency more than
25-30\% are used in the experiment.

\subsection{The contribution of $^{191}Os$}
The contribution of $^{191}Os$ appears, in particular, in the sum coincidence
spectrum as overlapping of full energy peaks related with cascade
transitions in $^{193}Os$ and $^{191}Os$. As can be seen from Table  3, such
overlapping affects 4 cascade intensity distributions measured in
$^{193}Os$. It should be noted that this effect was taken into consideration
only for final levels of cascades with $J_f\leq 5/2$ because, according to
all previous experiments, the intensity of cascades which include even
 though one quadrupole transition is considerably
less than that of cascades with two dipole transitions. 

The overlapping of peaks corresponding to different isotopes brings in the
necessity to remove well separated, intense cascades belonging to the
$^{191}Os$ isotope from Table~1 and correct the data in Table~2 for its
contribution. The correction has sense only if the cascades of $^{193}Os$
determine the major part of the area of a given doublet in the sum coincidence
spectrum. As a result, the experimentally  resolved cascades of $^{191}Os$
are removed from all 4 spectra.

 The removed cascades are attributed to $^{191}Os$
if within the limits of three standard errors of determination of the
intermediate level or $\gamma$-transition energy:

(a)~intermediate levels with a corresponding energy are not observed in
other 7 spectra of $^{193}Os$; but

(b)~the $\gamma$-transition with a close energy is observed in cascade
primary transitions of $^{191}Os$.

Of course, this procedure does not guarantee an absolute confidence of
results. It is, however, more suitable for determination of the level energies
than for the determination of decay modes of excited states. A number
of cascade transitions observed with a relatively large mean error of
determination of their energies ($\Delta E=0.36$ keV) does not allow one to
suggest a more reliable method to exclude cascades belonging to $^{191}Os$.
Moreover, presently available [15] information on thermal neutron radiative
capture spectra of $^{190,192}Os$ is considerably poorer than the data
obtained in the reported experiment and cannot be used to solve the problem
under consideration.

Correction of the total cascade intensities (Table~2) for the contribution of
$^{191}Os$ can be done in a simpler way. An analysis shows that the low-lying
levels of $^{191,193}Os$ with equal $J_f^{\pi}$ and the same structure are
populated by two-step cascades with approximately equal probabilities.
It should be noted that approximate equality of intensities is observed
also for the cascades terminating at the levels with different spins
$J=1/2$ or $3/2$ but equal parity.
Therefore, the ratio between the contributions of two
isotopes in the case of such $J_f^{\pi}$ was taken equal to the ratio
between the number of neutron captures in $^{190}Os$ and $^{192}Os$. Besides,
the intensity of cascades terminating at the $J_f^{\pi}=5/2^-$ level of
$^{193}Os$ was assumed to be two times less than the cascade intensity to the
$J=1/2, 3/2$ final levels of $^{191}Os$ under conditions of equality of
numbers of neutron captures in both nuclei.

The results of the procedure are taken into account in the data listed in
Table~2. Certainly, it is an approximate solution of the problem. Unresolved
doublets represent an insignificant part of the total cascade intensity,
however. Hence, one may expect negligible influence of the corresponding error
on physical results.

The maximum number of primary transitions belonging to $^{191}Os$ (but
entered in Table~1) can be estimated from frequency distribution of the
difference between the energies of the primary transitions in corresponding
spectroscopic data. Such frequency distribution for the 1 keV energy bins
in $^{191,193}Os$ is shown in Fig.~3. The enhanced (as compared with
neighboring intervals) frequency of observation of the primary transitions
with close energy can testify to admixture of the cascades of $^{191}Os$
in Table 1. Mistaken  isotopic identification is observed, probably, for
39(16) transitions from more than 500 primary transitions measured in two
isotopes. (Small but statistically significant deviation of the data given in
Fig.~3 from zero can be related to error in energy calibration for different
$Os$ isotopes). There is rather clear demonstration of reliability  of the
spectroscopic data obtained in our experiment.

\subsection{Comparison with a known decay scheme}
Investigation of cascade $\gamma$-decay of heavy compound nuclei is a
sensitive tool for obtaining spectroscopic information and reliable
establishing of a decay scheme up to the excitation energy of 3-4 MeV. The
observation confidence of nuclear excited states is mainly determined
by the intensity of populating cascades and depends weakly on the
excitation energy. For these reasons, the decay scheme of $^{193}Os$ above
$\simeq 1$ MeV established in our experiment seems to be more precise and
reliable than obtained earlier.

All observed by us levels of $^{193}Os$ with $J<5/2$ are listed in Table~1.
The known decay scheme of this nucleus includes only the part of possible
levels from those listed in Table~1. This is due to its construction mainly
on the basis of spectra of the primary $\gamma$-transitions following thermal
neutron radiative capture. These spectra, however, were measured at real
detection threshold  $L_c \simeq 5 \times 10^{-3}$ events per decay.
In our experiment there are observed the states of $^{193}Os$ populated by
the cascades with the sum intensity exceeding $2-4 \times 10^{-4}$ events
per decay. The energy of intermediate levels for the majority of these
cascades was determined according to [8] reliably enough, too. In the case of
low intensity cascades, quanta ordering was determined [7] only for part
of them. Some of the cascades were observed only in one of 11 distributions
and one cannot exclude that, possibly, some of the cascades listed in Table~1
can have  the low energy primary and high energy secondary transitions.

\section{Estimation of possible number of excited states below detection 
threshold}
At zero detection threshold $L_c=0$ decay scheme would include the main part
of excited states of the studied nucleus up to the excitation energy where
the mean spacing between the levels is comparable with the widths of peaks
(Fig.~2).

Of course, this is impossible. However, the most probable number of levels
populated and by the cascades with  $<I_{\gamma\gamma}> \leq L_c$ can be
estimated if it is taken into account that:

(a)~intensity of cascades is the random value with the corresponding average
and dispersion;

(b)~usually  $L_c$ is noticeably less than the mean intensity
$<I_{\gamma\gamma}>$ of cascades summed over their final levels (for a
given $E_i$).

Corresponding method of analysis is described in [16]. Its main idea is
that one can select such values of three parameters -- the set of the mean
intensities of cascades with $E1$ or $M1$ primary transitions (in reality -
sum of intensities for each intermediate level), the total number of
intermediate levels for a given interval of the excitation energy, and $L_c$
-- which provide the best reproduction of the cumulative sum of intensities
for $<I_{\gamma\gamma}>>L_c$. These parameters of approximation are
extrapolated over all region of possible values of cascade intensities on the
ground of the condition (b).

Experimental
cumulative distribution of $<I_{\gamma\gamma}>$ for some energy intervals
of their intermediate levels are compared with the results of the best
approximation in Fig.~4. If fluctuations of partial widths of the primary
transitions relative to their mean value are described by the Porter-Thomas
distribution [17] than the best values of the obtained parameters allow the
conclusion that:

1.~Table~1 contains more than 99\% of the total intensity of
cascades populating the levels with $E_i <3$ MeV and $E_f\leq 890$ keV.

2.~ The function approximating experimental cumulative sums in the studied
nucleus is identical in two possible variants:

(a)~or the intensities of cascades with the $E1$ and $M1$ primary transitions
and number of their intermediate levels practically are equal;

(b)~or the mean intensity of these cascades differs from that of cascades
with other type of the transitions by a factor of $\approx$ 100 and more and their
number equals the sum of the probable number of the intermediate levels
according to the variant (a).

3.~At the excitation energy above  $\simeq 2$ MeV the analysis [16] gives
considerably less number of levels than that predicted by the back-shifted
Fermi-gas model with the parameters taken from [18].

Available information on the level density is compared in Fig.~5.

It should be noted that a decrease in the model level density used in
calculation always leads to an increase in the intensity of the cascades
proceeding between the compound state and any low-lying level.
The use of the models of the radiative strength functions with the stronger
energy dependence also leads to increase in cascade intensity.
 As it is seen from the Table~2,
the calculated cascade intensity is considerably less than the experimental
values. This result confirms the conclusions of the analysis [16] that
the density of the levels excited in $^{193}Os$ is considerably less than
that predicted by the models which do not take into account the influence
on nucleon pairing at $E_{ex}>2-3$ MeV or underestimate it [19]. It should
be noted also that $^{193}Os$ is the first nucleus where the calculated
cascade intensities are stronger affected by choice the model of the
radiative strength functions than by choice of the level density model.
But on the whole, theoretical notions  of the level density and radiative
strength functions for $^{193}Os$ are to be developed completely in the same
direction as that earlier obtained [22] for other nuclei.

\section{Probable dominant component of the wave functions of the
intermediate levels of most intense cascades}
According to the modern theoretical notions, the wave function structure of
any excited states is determined by co-existence and interaction between the
fermion (quasiparticles) and boson (phonons) excitations. With the excitation
energy a nucleus transits from practically mono-component excitations of the
mentioned types to the mixed (quasiparticles $\otimes$ phonons) states with
rather different [23] degree of their fragmentation. This process should be
investigated in details but there is no adequate experimental methods to
study the structure of the wave functions above the excitation energy of 1-3
MeV.

Nevertheless, some information on the probable dominant components of wave
functions of heavy deformed nuclei can be obtained. The authors of [24]
suggested to search for the regularity in the excitation spectra of the
intermediate levels of most intense cascades by means of auto-correlation
analysis of the smoothed distributions of the sum cascade intensities
from Table~1. Intensities were smoothed by means of the Gaussian function:
$F(E)=\sum_{E}I_{\gamma\gamma} \times exp(-0.5(\Delta E/\sigma)^2)$.
The distribution of this type smoothed with the parameter
$\sigma=25$ keV is given in Fig.~6 and the values of the auto-correlation
function
\begin{equation}
A(T)=\sum_{E}F(E)\times F(E+T)\times F(E+2T).
\end{equation}
for different selection thresholds of intense cascades are shown in Fig.~7.
As it was shown in [25] such analysis cannot give unique value of the
equidistant period $T$ even for the simulated spectra (for example, for 25
``bands" consisting from 4 levels with slightly distorted equidistant period)
and provide estimation of the confidence level of the observed effect. In
principle, both problems can be solved in the experiments on the study of
the two-step cascades in different resonances of the same nucleus.
But some grounds to state that the regularity really exists can be obtained
from a comparison of the most probable equidistant periods in different
nuclei. The set of the probable equidistant periods obtained so far (Fig.~8)
allows an assumption that the $T$ value is approximately proportional to the
number of boson pairs of the unfilled nuclear shells. This allows one
to consider the effect at the level of working hypothesis.

The regularity in the excitation spectra testifies to the harmonic
nuclear vibrations. Thus, one can assume that the structure of the
intermediate levels of the studied cascades contains considerable components
of the rather weakly fragmented states like multi-quasiparticle excitations
$\otimes$ phonon or several phonons. This provides logical explanation of
serious decrease in the observed level density as compared with the
predictions of the non-interacting Fermi-gas model: nuclear excitation
energy concentrates on phonos but quasiparticles up to the
$\simeq 2$ MeV are populate weakly or very weakly due to insufficiency of
energy for breaking of paired nucleons.

\section{Influence of structure of final level on cascade intensity}
The structure of some levels of $^{193}Os$ is known: the ground state and
level $E_f$=72 keV are the members of the rotational band
$[Nn_z\Lambda]=[512]\hspace*{-4pt}\downarrow$, levels  $E_f$=41, 102, and,
probably, 295 keV belong to the band $[510]\hspace*{-4pt}\uparrow$.
It is seen from the Table~2 that the ratio between the experimental and
calculated cascades intensities amounts on the average to 1.9 and 2.7 in the
first and second cases, respectively. This discrepancy, as that observed
earlier in other even-odd compound nuclei, may be related [2] with the
influence of the structure of the cascade final level on mean probability
of the cascade.

\section{Conclusion}
Information on two-step $\gamma$-cascades for a number of nuclei from
the mass range $114\leq A\leq 200$ (see, for example, [26]) from thermal
neutron capture experiments forms a basis for study of the characteristics
of the $\gamma$-decay process.

The results indicate necessity to modify model notions of the
properties of the excited states of the heavy nuclei.
The obtained information demonstrates that more correct description of the
process under study requires more detailed accounting for co-existence and
interaction of superfluid and normal phases of nuclear matter by the
nuclear models. Achievement of complete correspondence between the observed
and calculated parameters of nuclear reactions, for instance, neutron-induced
reaction is impossible. This concerns, partially, the total radiative
widths of neutron resonances and $\gamma$-spectrum .
\\\\

This work was supported by GACR under contract No. 202/97/K038 and by RFBR
Grant No. 99-02-17863.

\newpage
\twocolumn
\parbox{170mm}
{\sl Table~1.\\ A list of absolute intensities (per $10^2$ decays),
$i_{\gamma\gamma}$, of measured two-step cascades and energies, $E_1$ and $E_2$, of
the cascade transitions,
 $E_i$ is the energy of the intermediate levels.}
\begin{center}
\begin{tabular}{|l|c|l|l|}  \hline
$E_1$,keV&  $E_i$, keV &$E_2$, keV &
$i_{\gamma\gamma}$\\ \hline
 5277.00&  306.90(10)&  306.90&   0.326(16)\\
                     &&  265.42&   2.209(52)\\
                     &&  204.17&   0.374(13)\\
 5033.00&  550.90(20)&  448.17&   0.032(4)\\
 5010.70&  573.20(20)&  266.12&   0.055(12)\\
 4996.30&  587.60(10)&  353.74&   0.041(6)\\
 4908.70&  675.20(10)&  633.72&   0.017(4)\\
                     &&  572.47&   0.013(4)\\
 4795.40&  788.50(10)&  788.50&   0.025(5)\\
 4694.50&  889.40(10)&  889.40&   1.031(19)\\
                     &&  847.92&   0.087(9)\\
                     &&  816.50&   0.165(8)\\
                     &&  786.67&   0.821(27)\\
                     &&  655.54&   0.183(16)\\
                     &&  582.32&   0.207(45)\\
 4617.00&  966.90(18)&  671.22&   0.014(4)\\
                     &&  567.88&   0.011(3)\\
 4530.20& 1053.70(20)& 1053.70&   0.312(8)\\
                     && 1012.22&   0.688(15)\\
                     &&  980.80&   0.065(5)\\
                     &&  950.97&   5.791(49)\\
                     &&  746.62&   0.500(38)\\
 4499.40& 1084.50(45)& 1043.02&   0.033(4)\\
                     &&  788.82&   0.070(9)\\
 4413.20& 1170.70(40)& 1170.70&   0.097(8)\\
                     && 1129.22&   0.806(24)\\
                     && 1097.80&   0.171(10)\\
                     && 1067.97&   0.226(10)\\
                     &&  936.84&   0.176(16)\\
                     &&  875.02&   0.161(12)\\
                     &&  863.62&   0.034(11)\\
                     &&  714.93&   0.030(6)\\
                     &&  461.50&   0.083(7)\\
 \hline \end{tabular} \end{center}
\newpage
\vspace*{9.8mm}
\begin{center}
\begin{tabular}{|l|c|l|l|}\hline
$E_1$,keV&  $E_i$, keV &$E_2$, keV &
$i_{\gamma\gamma}$\\ \hline
 4405.60& 1178.30(20)& 1178.30&   0.717(19)\\
                     && 1136.82&   0.029(6)\\
                     && 1105.40&   0.107(8)\\
                     && 1075.57&   0.342(11)\\
                     &&  871.22&   0.097(15)\\
                     &&  779.28&   0.020(7)\\
                     &&  722.53&   0.131(9)\\
 4398.50& 1185.40(20)& 1082.67&   0.023(6)\\
 4378.70& 1205.20(50)& 1163.72&   0.076(9)\\
                     && 1102.47&   0.097(6)\\
 4366.40& 1217.50(26)& 1217.50&   0.176(9)\\
                     && 1176.02&   0.593(21)\\
                     && 1144.60&   0.090(8)\\
                     && 1114.77&   0.433(13)\\
                     &&  921.82&   0.024(7)\\
                     &&  910.42&   0.066(12)\\
                     &&  818.48&   0.094(10)\\
                     &&  761.73&   0.072(6)\\
 4358.20& 1225.70(41)& 1184.22&   0.026(6)\\
                     && 1152.80&   0.020(6)\\
                     &&  769.93&   0.448(16)\\
 4339.30& 1244.60(39)& 1203.12&   0.015(6)\\
                     &&  788.83&   0.062(6)\\
 4316.60& 1267.30(27)& 1164.57&   0.093(10)\\
 4301.00& 1282.90(20)& 1282.90&   0.170(9)\\
                     && 1241.42&   0.100(6)\\
                     && 1210.00&   0.036(4)\\
                     && 1180.17&   0.068(8)\\
                     &&  987.22&   0.032(7)\\
                     &&  883.88&   0.119(14)\\
                     &&  573.70&   0.031(6)\\
 4295.30& 1288.60(10)& 1288.60&   0.630(18)\\
                     && 1247.12&   0.127(6)\\
 \hline \end{tabular} \end{center}
 \newpage
\vspace*{1mm}
{\footnotesize\hspace{35mm}
{\em   }\\}
\begin{center}
\begin{tabular}{|l|c|l|l|}  \hline
$E_1$,keV&  $E_i$, keV &$E_2$, keV &
$i_{\gamma\gamma}$\\ \hline
                     && 1215.70&   0.014(4)\\
                     && 1185.87&   0.040(8)\\
                     &&  992.92&   0.064(7)\\
                     &&  889.58&   0.081(11)\\
                     &&  579.40&   0.063(6)\\
                     &&  399.12&   0.115(12)\\
 4250.40& 1333.50(30)& 1333.50&   0.060(5)\\
                     && 1292.02&   0.083(6)\\
                     && 1260.60&   0.017(4)\\
                     && 1230.77&   0.448(17)\\
                     && 1099.64&   0.112(14)\\
                     && 1037.82&   0.021(7)\\
                     &&  934.48&   0.026(9)\\
                     &&  877.73&   0.029(12)\\
 4224.30& 1359.60(30)& 1359.60&   0.045(5)\\
                     && 1318.12&   0.059(5)\\
                     && 1286.70&   0.035(4)\\
                     && 1256.87&   0.108(10)\\
                     && 1125.74&   0.851(37)\\
                     && 1063.92&   0.047(5)\\
                     && 1052.52&   0.065(6)\\
                     &&  960.58&   0.144(15)\\
                     &&  903.83&   0.095(14)\\
 4200.30& 1383.60(30)& 1076.52&   0.029(6)\\
 4197.90& 1386.00(80)& 1386.00&   0.754(18)\\
                     && 1344.52&   0.140(9)\\
                     && 1313.10&   0.029(5)\\
                     && 1283.27&   0.019(4)\\
                     && 1090.32&   0.027(5)\\
                     && 1078.92&   0.096(7)\\
 4185.70& 1398.20(36)& 1398.20&   0.069(8)\\
                     && 1325.30&   0.032(5)\\
 \hline \end{tabular} \end{center}
 \newpage
{\footnotesize\hspace{35mm}
{\em Table 1 (continued)}\\}
\vspace*{1mm}
\begin{center}
\begin{tabular}{|l|c|l|l|}  \hline
$E_1$,keV&  $E_i$, keV &$E_2$, keV &
$i_{\gamma\gamma}$\\ \hline
 4183.90& 1400.00(30)& 1400.00&   0.034(6)\\
                     && 1358.52&   0.017(6)\\
                     && 1297.27&   0.199(8)\\
                     && 1092.92&   0.043(6)\\
 4165.90& 1418.00(26)& 1418.00&   0.066(6)\\
                     && 1376.52&   0.016(5)\\
                     && 1345.10&   0.045(5)\\
                     && 1122.32&   0.022(5)\\
 4149.90& 1434.00(70)& 1434.00&   0.014(5)\\
 4137.40& 1446.50(57)& 1373.60&   0.112(6)\\
 4086.50& 1497.40(45)& 1497.40&   0.018(5)\\
 4079.80& 1504.10(41)& 1504.10&   0.024(5)\\
                     && 1401.37&   0.025(6)\\
                     && 1270.24&   0.070(6)\\
                     && 1048.33&   0.038(9)\\
 4068.30& 1515.60(100)& 1515.60&   0.021(5)\\
                     && 1474.12&   0.051(6)\\
                     && 1412.87&   0.085(8)\\
                     && 1219.92&   0.054(8)\\
                     && 1208.52&   0.200(11)\\
                     && 1116.58&   0.083(11)\\
                     && 1059.83&   0.060(10)\\
 4060.40& 1523.50(40)& 1523.50&   0.069(6)\\
                     && 1482.02&   0.067(6)\\
                     && 1420.77&   0.207(11)\\
                     && 1216.42&   0.207(11)\\
 4053.60& 1530.30(29)& 1488.82&   0.248(10)\\
                     && 1427.57&   0.327(13)\\
                     && 1234.62&   0.037(8)\\
                     && 1223.22&   0.062(6)\\
                     && 1131.28&   0.035(11)\\
 4028.10& 1555.80(20)& 1514.32&   0.060(6)\\
 \hline \end{tabular} \end{center}
 \newpage
{\footnotesize\hspace{35mm}
{\em     }\\}
\begin{center}
\begin{tabular}{|l|c|l|l|}  \hline
$E_1$,keV&  $E_i$, keV &$E_2$, keV &
$i_{\gamma\gamma}$\\ \hline
                     && 1248.72&   0.035(6)\\
 3993.00& 1590.90(80)& 1590.90&   0.222(11)\\
                     && 1549.42&   0.040(5)\\
                     && 1518.00&   0.049(5)\\
                     && 1488.17&   0.087(8)\\
                     && 1295.22&   0.046(6)\\
                     && 1283.82&   0.139(12)\\
 3980.70& 1603.20(32)& 1603.20&   0.046(6)\\
                     && 1561.72&   0.671(17)\\
                     && 1530.30&   0.023(5)\\
                     && 1500.47&   0.610(17)\\
                     && 1369.34&   0.029(6)\\
                     && 1307.52&   0.178(10)\\
                     && 1296.12&   0.272(16)\\
                     && 1204.18&   0.212(16)\\
                     && 1147.43&   0.155(14)\\
 3923.60& 1660.30(30)& 1618.82&   0.017(5)\\
 3903.60& 1680.30(42)& 1373.22&   0.024(8)\\
 3900.60& 1683.30(17)& 1683.30&   0.091(9)\\
                     && 1641.82&   0.350(12)\\
                     && 1610.40&   0.021(5)\\
                     && 1580.57&   0.445(15)\\
                     && 1449.44&   0.250(12)\\
                     && 1387.62&   0.138(9)\\
                     && 1376.22&   0.072(9)\\
                     &&  974.10&   0.042(6)\\
 3861.40& 1722.50(30)& 1722.50&   0.019(5)\\
                     && 1619.77&   0.017(6)\\
                     && 1426.82&   0.018(6)\\
 3852.30& 1731.60(33)&  842.12&   0.058(9)\\
 3846.30& 1737.60(60)& 1634.87&   0.025(6)\\
 3839.00& 1744.90(90)& 1672.00&   0.016(5)\\
                     && 1289.13&   0.016(5)\\
 \hline \end{tabular} \end{center}
 \newpage
{\footnotesize\hspace{35mm}
{\em Table 1 (continued)}\\}
\begin{center}
\begin{tabular}{|l|c|l|l|}  \hline
$E_1$,keV&  $E_i$, keV &$E_2$, keV &
$i_{\gamma\gamma}$\\ \hline
 3829.70& 1754.20(110)& 1754.20&   0.016(5)\\
                     && 1520.34&   0.016(4)\\
 3823.50& 1760.40(20)& 1657.67&   0.047(8)\\
                     && 1526.54&   0.014(4)\\
 3818.80& 1765.10(85)& 1723.62&   0.074(6)\\
                     && 1692.20&   0.027(5)\\
                     && 1662.37&   0.108(8)\\
                     && 1531.24&   0.102(6)\\
                     && 1458.02&   0.117(8)\\
                     && 1366.08&   0.056(5)\\
                     && 1309.33&   0.048(5)\\
                     && 1055.90&   0.025(6)\\
 3800.10& 1783.80(90)& 1783.80&   0.065(6)\\
                     && 1742.32&   0.112(7)\\
                     && 1710.90&   0.033(5)\\
                     && 1549.94&   0.129(7)\\
                     && 1488.12&   0.025(5)\\
                     && 1476.72&   0.035(6)\\
                     && 1384.78&   0.035(5)\\
                     && 1328.03&   0.030(5)\\
 3788.10& 1795.80(43)& 1693.07&   0.021(6)\\
                     && 1561.94&   0.019(4)\\
 3785.00& 1798.90(50)& 1565.04&   0.030(4)\\
 3781.90& 1802.00(75)& 1760.52&   0.022(5)\\
 3778.80& 1805.10(34)& 1805.10&   0.026(5)\\
 3757.20& 1826.70(90)& 1826.70&   0.018(5)\\
                     && 1753.80&   0.014(5)\\
 3752.80& 1831.10(28)& 1831.10&   0.029(5)\\
                     && 1789.62&   0.037(5)\\
                     && 1758.20&   0.032(5)\\
                     && 1597.24&   0.037(4)\\
                     && 1535.42&   0.027(5)\\
                     && 1524.02&   0.018(6)\\
 \hline \end{tabular} \end{center}
 \newpage
{\footnotesize\hspace{35mm}
{\em    }\\}
\begin{center}
\begin{tabular}{|l|c|l|l|}  \hline
$E_1$,keV&  $E_i$, keV &$E_2$, keV &
$i_{\gamma\gamma}$\\ \hline
                     && 1432.08&   0.020(4)\\
                     && 1375.33&   0.022(5)\\
 3745.60& 1838.30(20)& 1765.40&   0.021(5)\\
                     && 1735.57&   0.025(6)\\
                     && 1604.44&   0.073(5)\\
                     && 1531.22&   0.399(15)\\
                     && 1129.10&   0.051(9)\\
 3736.80& 1847.10(52)& 1847.10&   0.026(6)\\
                     && 1805.62&   0.065(6)\\
                     && 1774.20&   0.024(5)\\
                     && 1744.37&   0.080(8)\\
                     && 1551.42&   0.019(5)\\
                     && 1540.02&   0.028(6)\\
                     && 1448.08&   0.073(5)\\
                     && 1137.90&   0.047(9)\\
 3730.30& 1853.60(92)& 1619.74&   0.016(4)\\
 3721.20& 1862.70(80)& 1759.97&   0.029(6)\\
 3709.30& 1874.60(45)& 1874.60&   0.031(10)\\
 3695.00& 1888.90(84)& 1816.00&   0.029(5)\\
                     && 1786.17&   0.032(8)\\
                     && 1489.88&   0.024(5)\\
                     && 1179.70&   0.037(9)\\
 3691.30& 1892.60(27)& 1892.60&   0.025(10)\\
                     && 1851.12&   0.040(5)\\
                     && 1819.70&   0.040(5)\\
                     && 1436.83&   0.021(6)\\
 3675.30& 1908.60(35)& 1867.12&   0.015(5)\\
                     && 1509.58&   0.032(6)\\
                     && 1452.83&   0.025(6)\\
 3668.60& 1915.30(43)& 1915.30&   0.029(10)\\
                     && 1873.82&   0.411(12)\\
                     && 1812.57&   0.030(6)\\
                     && 1681.44&   0.294(10)\\
 \hline \end{tabular} \end{center}
 \newpage
{\footnotesize\hspace{35mm}{\em Table 1 (continued)}\\}
\begin{center}
\begin{tabular}{|l|c|l|l|}  \hline
$E_1$,keV&  $E_i$, keV &$E_2$, keV &
$i_{\gamma\gamma}$\\ \hline
                     && 1608.22&   0.416(16)\\
                     && 1516.28&   0.019(6)\\
                     && 1459.53&   0.022(6)\\
 3662.70& 1921.20(30)& 1848.30&   0.013(5)\\
                     && 1818.47&   0.030(6)\\
 3651.80& 1932.10(60)& 1890.62&   0.028(5)\\
                     && 1859.20&   0.089(5)\\
 3648.80& 1935.10(16)& 1935.10&   0.023(10)\\
                     && 1832.37&   0.082(6)\\
                     && 1536.08&   0.023(5)\\
 3645.30& 1938.60(44)& 1938.60&   0.321(24)\\
                     && 1897.12&   0.048(5)\\
 3634.90& 1949.00(42)& 1641.92&   0.018(6)\\
 3629.10& 1954.80(20)& 1852.07&   0.355(11)\\
                     && 1720.94&   0.022(4)\\
                     && 1647.72&   0.016(6)\\
                     && 1499.03&   0.017(5)\\
 3606.50& 1977.40(90)& 1935.92&   0.037(5)\\
                     && 1743.54&   0.030(4)\\
                     && 1670.32&   0.107(8)\\
                     && 1578.38&   0.021(5)\\
                     && 1087.92&   0.054(10)\\
 3600.50& 1983.40(22)& 1983.40&   0.160(16)\\
                     && 1910.50&   0.067(5)\\
                     && 1880.67&   0.070(6)\\
                     && 1749.54&   0.015(4)\\
                     && 1687.72&   0.020(6)\\
 3594.10& 1989.80(60)& 1755.94&   0.020(4)\\
 3581.80& 2002.10(28)& 1960.62&   0.210(11)\\
                     && 1929.20&   0.087(6)\\
                     && 1768.24&   0.048(5)\\
                     && 1706.42&   0.039(6)\\
                     && 1695.02&   0.040(6)\\
 \hline \end{tabular} \end{center}
 \newpage
{\footnotesize\hspace{35mm}
{\em     }\\}
\begin{center}
\begin{tabular}{|l|c|l|l|}  \hline
$E_1$,keV&  $E_i$, keV &$E_2$, keV &
$i_{\gamma\gamma}$\\ \hline
                     && 1603.08&   0.025(6)\\
 3570.30& 2013.60(37)& 1972.12&   0.123(9)\\
                     && 1910.87&   0.032(6)\\
                     && 1779.74&   0.015(4)\\
                     && 1706.52&   0.051(6)\\
                     && 1557.83&   0.017(6)\\
                     && 1124.12&   0.094(10)\\
 3563.10& 2020.80(80)& 1786.94&   0.022(4)\\
                     && 1713.72&   0.021(6)\\
 3559.60& 2024.30(55)& 1982.82&   0.027(6)\\
 3546.50& 2037.40(80)& 1803.54&   0.030(5)\\
                     && 1741.72&   0.022(6)\\
                     && 1730.32&   0.018(5)\\
                     && 1581.63&   0.030(6)\\
 3544.00& 2039.90(70)& 2039.90&   0.043(5)\\
                     && 1732.82&   0.021(5)\\
 3535.80& 2048.10(30)& 2048.10&   0.141(8)\\
                     && 1945.37&   0.049(8)\\
                     && 1814.24&   0.049(6)\\
                     && 1741.02&   0.040(5)\\
 3533.10& 2050.80(80)& 1816.94&   0.023(5)\\
                     && 1755.12&   0.046(6)\\
                     && 1651.78&   0.046(5)\\
 3530.40& 2053.50(85)& 1950.77&   0.080(10)\\
 3524.20& 2059.70(20)& 2018.22&   0.018(6)\\
 3519.80& 2064.10(16)& 2064.10&   0.039(5)\\
                     && 2022.62&   0.051(7)\\
                     && 1991.20&   0.076(6)\\
                     && 1961.37&   0.097(10)\\
                     && 1757.02&   0.053(5)\\
                     && 1665.08&   0.059(6)\\
                     && 1608.33&   0.049(6)\\
 \hline \end{tabular} \end{center}
 \newpage
{\footnotesize\hspace{35mm}{\em Table 1 (continued)}\\}
\begin{center}
\begin{tabular}{|l|c|l|l|}  \hline
$E_1$,keV&  $E_i$, keV &$E_2$, keV &
$i_{\gamma\gamma}$\\ \hline
                    && 1354.90&   0.019(7)\\
 3516.30& 2067.60(8)& 2067.60&   0.030(5)\\
                     && 1833.74&   0.082(6)\\
                     && 1760.52&   0.043(5)\\
                     && 1178.12&   0.052(8)\\
 3505.60& 2078.30(50)& 2078.30&   0.039(5)\\
                     && 2036.82&   0.295(12)\\
                     && 2005.40&   0.031(5)\\
                     && 1844.44&   0.052(5)\\
                     && 1782.62&   0.020(6)\\
                     && 1771.22&   0.115(7)\\
                     && 1679.28&   0.027(5)\\
                     && 1622.53&   0.043(6)\\
                     && 1369.10&   0.026(7)\\
                     && 1188.82&   0.032(8)\\
 3502.80& 2081.10(46)& 2008.20&   0.054(5)\\
                     && 1978.37&   0.042(8)\\
                     && 1785.42&   0.027(6)\\
                     && 1774.02&   0.035(5)\\
                     && 1682.08&   0.052(6)\\
                     && 1625.33&   0.025(6)\\
 3491.00& 2092.90(20)& 2092.90&   0.020(5)\\
                     && 2020.00&   0.022(5)\\
                     && 1990.17&   0.038(8)\\
                     && 1859.04&   0.032(5)\\
                     && 1797.22&   0.038(6)\\
                     && 1693.88&   0.111(7)\\
                     && 1203.42&   0.028(8)\\
 3485.90& 2098.00(48)& 2098.00&   0.024(5)\\
                     && 2056.52&   0.021(6)\\
                     && 1995.27&   0.021(8)\\
                     && 1864.14&   0.045(5)\\
 \hline \end{tabular} \end{center}
 \newpage
{\footnotesize\hspace{35mm}
{\em      }\\}
\begin{center}
\begin{tabular}{|l|c|l|l|}  \hline
$E_1$,keV&  $E_i$, keV &$E_2$, keV &
$i_{\gamma\gamma}$\\ \hline
                     && 1802.32&   0.023(6)\\
                     && 1790.92&   0.020(5)\\
                     && 1388.80&   0.047(7)\\
                     && 1208.52&   0.045(8)\\
 3480.50& 2103.40(40)& 2061.92&   0.018(6)\\
                     && 1807.72&   0.018(6)\\
 3475.80& 2108.10(80)& 1874.24&   0.022(6)\\
 3472.20& 2111.70(70)& 2111.70&   0.024(5)\\
                     && 1222.22&   0.034(8)\\
 3468.00& 2115.90(50)& 2074.42&   0.032(6)\\
 3459.80& 2124.10(32)& 2021.37&   0.046(8)\\
                     && 1890.24&   0.022(4)\\
 3457.50& 2126.40(32)& 2084.92&   0.032(6)\\
                     && 1819.32&   0.032(5)\\
 3450.90& 2133.00(67)& 2133.00&   0.030(5)\\
                     && 2030.27&   0.017(6)\\
                     && 1899.14&   0.054(7)\\
                     && 1837.32&   0.037(5)\\
                     && 1733.98&   0.037(5)\\
                     && 1677.23&   0.054(6)\\
                     && 1243.52&   0.168(9)\\
 3449.70& 2134.20(40)& 2134.20&   0.021(5)\\
                     && 1900.34&   0.048(6)\\
                     && 1827.12&   0.086(6)\\
 3440.40& 2143.50(42)& 2040.77&   0.015(6)\\
 3433.30& 2150.60(52)& 2109.12&   0.018(7)\\
                     && 1854.92&   0.023(5)\\
 3430.10& 2153.80(60)& 2051.07&   0.061(8)\\
                     && 1858.12&   0.039(5)\\
 3426.80& 2157.10(10)& 2157.10&   0.030(5)\\
                     && 1850.02&   0.028(5)\\
 3420.20& 2163.70(50)& 2090.80&   0.042(5)\\
                     && 1929.84&   0.026(4)\\
 \hline \end{tabular} \end{center}
 \newpage
{\footnotesize\hspace{35mm}
{\em Table 1 (continued)}\\}
\begin{center}
\begin{tabular}{|l|c|l|l|}  \hline
$E_1$,keV&  $E_i$, keV &$E_2$, keV &
$i_{\gamma\gamma}$\\ \hline
                     && 1707.93&   0.021(6)\\
 3415.20& 2168.70(40)& 2168.70&   0.111(8)\\
                     && 2127.22&   0.187(11)\\
                     && 2065.97&   0.017(6)\\
                     && 1873.02&   0.023(5)\\
                     && 1861.62&   0.026(5)\\
                     && 1769.68&   0.049(5)\\
 3405.80& 2178.10(10)& 1288.62&   0.042(11)\\
 3402.60& 2181.30(80)& 2139.82&   0.056(7)\\
                     && 2108.40&   0.019(5)\\
 3398.50& 2185.40(50)& 2143.92&   0.017(6)\\
                     && 2082.67&   0.146(10)\\
                     && 1878.32&   0.023(11)\\
                     && 1295.92&   0.048(11)\\
 3391.50& 2192.40(22)& 2150.92&   0.063(7)\\
                     && 2089.67&   0.078(8)\\
                     && 1483.20&   0.029(8)\\
 3388.90& 2195.00(31)& 2092.27&   0.046(8)\\
                     && 1887.92&   0.216(19)\\
                     && 1795.98&   0.016(5)\\
                     && 1305.52&   0.082(11)\\
 3378.80& 2205.10(35)& 2102.37&   0.040(6)\\
                     && 1806.08&   0.018(5)\\
 3365.30& 2218.60(25)& 2218.60&   0.030(5)\\
                     && 1984.74&   0.073(6)\\
                     && 1922.92&   0.033(5)\\
                     && 1819.58&   0.014(5)\\
 3361.90& 2222.00(26)& 2222.00&   0.014(5)\\
                     && 1988.14&   0.024(5)\\
 3358.80& 2225.10(30)& 2183.62&   0.251(12)\\
                     && 1929.42&   0.026(5)\\
 3353.30& 2230.60(28)& 2230.60&   0.023(5)\\
                     && 2127.87&   0.042(6)\\
 \hline \end{tabular} \end{center}
 \newpage
{\footnotesize\hspace{35mm}
{\em     }\\}
\begin{center}
\begin{tabular}{|l|c|l|l|}  \hline
$E_1$,keV&  $E_i$, keV &$E_2$, keV &
$i_{\gamma\gamma}$\\ \hline
 3349.30& 2234.60(70)& 2234.60&   0.021(5)\\
 3344.00& 2239.90(50)& 2198.42&   0.021(6)\\
 3337.60& 2246.30(70)& 2204.82&   0.029(6)\\
 3334.80& 2249.10(20)& 2176.20&   0.152(9)\\
                     && 2146.37&   0.067(8)\\
                     && 1942.02&   0.380(25)\\
                     && 1850.08&   0.088(7)\\
                     && 1359.62&   0.058(11)\\
 3334.70& 2249.20(95)& 1540.00&   0.061(12)\\
 3333.00& 2250.90(70)& 2250.90&   0.602(26)\\
                     && 2209.42&   0.028(6)\\
 3328.00& 2255.90(20)& 2153.17&   0.021(6)\\
                     && 1948.82&   0.031(11)\\
 3325.50& 2258.40(40)& 2258.40&   0.024(9)\\
                     && 1859.38&   0.022(5)\\
 3305.20& 2278.70(20)& 2205.80&   0.028(5)\\
                     && 2175.97&   0.021(6)\\
                     && 1983.02&   0.020(6)\\
 3298.50& 2285.40(40)& 2182.67&   0.055(6)\\
 3293.40& 2290.50(30)& 2290.50&   0.020(8)\\
                     && 2249.02&   0.033(5)\\
 3289.60& 2294.30(31)& 2294.30&   0.021(8)\\
 3286.60& 2297.30(60)& 2255.82&   0.032(5)\\
                     && 2224.40&   0.033(5)\\
                     && 2194.57&   0.017(6)\\
 3273.90& 2310.00(15)& 2310.00&   0.052(14)\\
                     && 2268.52&   0.045(5)\\
                     && 2076.14&   0.035(16)\\
 3268.00& 2315.90(78)& 2315.90&   0.026(9)\\
                     && 2274.42&   0.056(5)\\
                     && 2020.22&   0.022(6)\\
                     && 2008.82&   0.020(8)\\
 \hline \end{tabular} \end{center}
 \newpage
{\footnotesize\hspace{35mm}
{\em Table 1 (continued)}\\}
\begin{center}
\begin{tabular}{|l|c|l|l|}  \hline
$E_1$,keV&  $E_i$, keV &$E_2$, keV &
$i_{\gamma\gamma}$\\ \hline
                     && 1860.13&   0.064(8)\\
 3263.40& 2320.50(27)& 2320.50&   0.107(13)\\
                     && 2217.77&   0.076(6)\\
                     && 2086.64&   0.078(7)\\
                     && 2024.82&   0.020(6)\\
 3257.80& 2326.10(90)& 2326.10&   0.066(10)\\
                     && 2284.62&   0.048(5)\\
                     && 2223.37&   0.291(11)\\
                     && 2030.42&   0.027(6)\\
 3251.30& 2332.60(29)& 2291.12&   0.016(5)\\
                     && 2229.87&   0.017(6)\\
 3243.80& 2340.10(42)& 2298.62&   0.026(5)\\
 3241.00& 2342.90(80)& 2301.42&   0.029(5)\\
                     && 2240.17&   0.055(6)\\
 3235.90& 2348.00(24)& 2348.00&   0.032(5)\\
                     && 2306.52&   0.110(6)\\
                     && 2245.27&   0.055(6)\\
                     && 2114.14&   0.029(6)\\
 3233.50& 2350.40(40)& 2043.32&   0.019(8)\\
 3223.00& 2360.90(75)& 2127.04&   0.021(6)\\
 3219.70& 2364.20(37)& 2364.20&   0.024(5)\\
                     && 2261.47&   0.025(6)\\
 3215.90& 2368.00(29)& 2326.52&   0.052(6)\\
                     && 2265.27&   0.093(8)\\
                     && 2134.14&   0.172(11)\\
                     && 2060.92&   0.019(8)\\
                     && 1478.52&   0.064(8)\\
 3210.80& 2373.10(34)& 2373.10&   0.016(5)\\
 3202.90& 2381.00(27)& 2381.00&   0.043(5)\\
                     && 2308.10&   0.019(5)\\
                     && 2073.92&   0.020(8)\\
                     && 1491.52&   0.033(8)\\
 \hline \end{tabular} \end{center}
 \newpage
{\footnotesize\hspace{35mm}
{\em    }\\}
\begin{center}
\begin{tabular}{|l|c|l|l|}  \hline
$E_1$,keV&  $E_i$, keV &$E_2$, keV &
$i_{\gamma\gamma}$\\ \hline
 3194.80& 2389.10(43)& 2389.10&   0.153(10)\\
                     && 2347.62&   0.026(6)\\
                     && 2286.37&   0.055(6)\\
                     && 2155.24&   0.020(5)\\
                     && 2093.42&   0.023(6)\\
                     && 1990.08&   0.022(9)\\
                     && 1933.33&   0.028(10)\\
                     && 1499.62&   0.065(8)\\
 3187.60& 2396.30(31)& 2396.30&   0.046(5)\\
 3176.90& 2407.00(47)& 2407.00&   0.024(5)\\
                     && 2365.52&   0.072(6)\\
                     && 2304.27&   0.080(8)\\
                     && 2111.32&   0.041(7)\\
                     && 2007.98&   0.031(9)\\
 3169.90& 2414.00(85)& 2372.52&   0.021(6)\\
 3162.90& 2421.00(27)& 2379.52&   0.070(7)\\
                     && 2021.98&   0.017(5)\\
                     && 1531.52&   0.042(9)\\
 3157.10& 2426.80(40)& 2426.80&   0.055(6)\\
                     && 2324.07&   0.036(6)\\
                     && 2192.94&   0.017(5)\\
 3152.60& 2431.30(80)& 2431.30&   0.047(6)\\
                     && 2328.57&   0.093(10)\\
 3151.10& 2432.80(40)& 2432.80&   0.019(5)\\
                     && 2391.32&   0.155(9)\\
                     && 2330.07&   0.019(8)\\
                     && 2125.72&   0.037(9)\\
                     && 2033.78&   0.019(5)\\
 3146.20& 2437.70(40)& 2396.22&   0.017(6)\\
 3141.40& 2442.50(52)& 2442.50&   0.032(5)\\
 3136.90& 2447.00(100)& 2405.52&   0.051(6)\\
                     && 2151.32&   0.019(6)\\
 \hline \end{tabular} \end{center}
 \newpage
{\footnotesize\hspace{35mm}
{\em Table 1 (continued)}\\}
\begin{center}
\begin{tabular}{|l|c|l|l|}  \hline
$E_1$,keV&  $E_i$, keV &$E_2$, keV &
$i_{\gamma\gamma}$\\ \hline
3133.80& 2450.10(55)& 2377.20&   0.015(5)\\
                     && 2347.37&   0.019(6)\\
 3125.40& 2458.50(20)& 2355.77&   0.027(6)\\
                     && 2151.42&   0.017(6)\\
 3122.20& 2461.70(50)& 2461.70&   0.027(5)\\
                     && 2227.84&   0.110(7)\\
                     && 2154.62&   0.031(6)\\
 3116.20& 2467.70(50)& 2426.22&   0.052(6)\\
                     && 2364.97&   0.021(6)\\
                     && 2233.84&   0.022(5)\\
                     && 2160.62&   0.019(6)\\
                     && 2068.68&   0.020(5)\\
                     && 2011.93&   0.025(10)\\
 3113.50& 2470.40(29)& 2470.40&   0.025(8)\\
                     && 2163.32&   0.025(6)\\
 3099.60& 2484.30(20)& 2442.82&   0.057(6)\\
                     && 2381.57&   0.023(6)\\
                     && 2250.44&   0.035(5)\\
 3097.20& 2486.70(65)& 2179.62&   0.016(6)\\
 3094.30& 2489.60(25)& 2489.60&   0.044(6)\\
                     && 2386.87&   0.021(8)\\
                     && 2182.52&   0.024(6)\\
                     && 2090.58&   0.033(5)\\
 3088.90& 2495.00(70)& 2422.10&   0.070(5)\\
                     && 2261.14&   0.015(5)\\
                     && 2187.92&   0.024(5)\\
                     && 2039.23&   0.063(7)\\
 3084.20& 2499.70(24)& 2499.70&   0.027(6)\\
                     && 2458.22&   0.077(15)\\
                     && 2426.80&   0.027(5)\\
                     && 2396.97&   0.021(8)\\
                     && 2265.84&   0.025(5)\\
 \hline \end{tabular} \end{center}
 \newpage
{\footnotesize\hspace{35mm}
{\em    }\\}
\begin{center}
\begin{tabular}{|l|c|l|l|}  \hline
$E_1$,keV&  $E_i$, keV &$E_2$, keV &
$i_{\gamma\gamma}$\\ \hline
                    && 2204.02&   0.051(6)\\
 3080.40& 2503.50(32)& 2503.50&   0.019(6)\\
 3077.60& 2506.30(100)& 2433.40&   0.045(5)\\
 3075.60& 2508.30(25)& 2508.30&   0.083(8)\\
                     && 2466.82&   0.034(7)\\
                     && 2405.57&   0.027(8)\\
                     && 2274.44&   0.073(6)\\
                     && 2201.22&   0.022(5)\\
                     && 2109.28&   0.020(5)\\
                     && 2052.53&   0.044(6)\\
 3072.10& 2511.80(70)& 2511.80&   0.031(6)\\
                     && 2438.90&   0.040(5)\\
                     && 2409.07&   0.070(10)\\
                     && 2204.72&   0.023(6)\\
                     && 2112.78&   0.086(6)\\
                     && 2056.03&   0.028(6)\\
 3069.80& 2514.10(6)& 2514.10&   0.026(6)\\
                     && 2472.62&   0.023(7)\\
                     && 2280.24&   0.039(5)\\
                     && 2218.42&   0.020(6)\\
                     && 2207.02&   0.027(6)\\
 3064.70& 2519.20(50)& 2519.20&   0.032(6)\\
                     && 2477.72&   0.032(7)\\
                     && 2446.30&   0.018(5)\\
                     && 2416.47&   0.034(8)\\
                     && 2063.43&   0.035(6)\\
 3055.50& 2528.40(50)& 2528.40&   0.015(6)\\
                     && 2455.50&   0.030(5)\\
                     && 2425.67&   0.019(8)\\
                     && 2072.63&   0.026(6)\\
 3053.00& 2530.90(40)& 2223.82&   0.021(6)\\
 \hline \end{tabular} \end{center}
 \newpage
{\footnotesize\hspace{35mm}
{\em Table 1 (continued)}\\}
\begin{center}
\begin{tabular}{|l|c|l|l|}  \hline
$E_1$,keV&  $E_i$, keV &$E_2$, keV &
$i_{\gamma\gamma}$\\ \hline
 3050.20& 2533.70(70)& 2533.70&   0.105(9)\\
                     && 2299.84&   0.058(5)\\
 3042.10& 2541.80(40)& 2541.80&   0.015(6)\\
                     && 2468.90&   0.022(6)\\
                     && 2439.07&   0.076(10)\\
                     && 2307.94&   0.052(5)\\
                     && 2234.72&   0.051(6)\\
                     && 2086.03&   0.051(6)\\
 3035.70& 2548.20(60)& 2445.47&   0.044(8)\\
                     && 2314.34&   0.093(6)\\
                     && 2252.52&   0.029(7)\\
 3032.60& 2551.30(90)& 2551.30&   0.144(10)\\
 3029.30& 2554.60(80)& 2554.60&   0.016(6)\\
                     && 2247.52&   0.019(6)\\
 3025.80& 2558.10(44)& 2516.62&   0.081(7)\\
                     && 2251.02&   0.022(6)\\
 3023.50& 2560.40(30)& 2487.50&   0.030(5)\\
                     && 2457.67&   0.030(8)\\
                     && 2326.54&   0.035(5)\\
                     && 2253.32&   0.040(6)\\
 3016.80& 2567.10(15)& 2567.10&   0.070(9)\\
                     && 2525.62&   0.040(6)\\
                     && 2494.20&   0.057(6)\\
                     && 2260.02&   0.042(6)\\
                     && 2168.08&   0.034(6)\\
                     && 2111.33&   0.047(6)\\
 3005.90& 2578.00(50)& 2344.14&   0.027(9)\\
 3003.80& 2580.10(34)& 2538.62&   0.032(6)\\
                     && 2507.20&   0.045(5)\\
                     && 2477.37&   0.158(10)\\
                     && 2346.24&   0.052(9)\\
 \hline \end{tabular} \end{center}
 \newpage
{\footnotesize\hspace{35mm}
{\em       }\\}
\begin{center}
\begin{tabular}{|l|c|l|l|}  \hline
$E_1$,keV&  $E_i$, keV &$E_2$, keV &
$i_{\gamma\gamma}$\\ \hline
                    && 2284.42&   0.024(6)\\
                     && 2124.33&   0.037(6)\\
 2998.90& 2585.00(85)& 2585.00&   0.018(6)\\
                     && 2543.52&   0.030(6)\\
 2986.50& 2597.40(30)& 2597.40&   0.067(9)\\
                     && 2363.54&   0.019(5)\\
 2981.10& 2602.80(80)& 2602.80&   0.021(6)\\
                     && 2561.32&   0.029(7)\\
                     && 2529.90&   0.043(5)\\
                     && 2203.78&   0.028(6)\\
                     && 2147.03&   0.022(6)\\
 2977.00& 2606.90(32)& 2504.17&   0.023(6)\\
 2972.60& 2611.30(25)& 2611.30&   0.020(6)\\
                     && 2569.82&   0.233(13)\\
                     && 2508.57&   0.049(6)\\
                     && 2315.62&   0.048(6)\\
                     && 1902.10&   0.030(8)\\
 2969.20& 2614.70(14)& 2380.84&   0.028(5)\\
 2954.60& 2629.30(24)& 2526.57&   0.044(6)\\
                     && 2322.22&   0.015(6)\\
 2951.60& 2632.30(10)& 2632.30&   0.049(8)\\
                     && 2590.82&   0.041(7)\\
                     && 2529.57&   0.030(6)\\
 2946.10& 2637.80(30)& 2637.80&   0.018(6)\\
                     && 2596.32&   0.039(6)\\
                     && 2535.07&   0.015(6)\\
 2927.30& 2656.60(42)& 2656.60&   0.043(6)\\
                     && 2615.12&   0.057(7)\\
                     && 2553.87&   0.040(8)\\
 2922.10& 2661.80(35)& 2354.72&   0.020(6)\\
 2912.50& 2671.40(30)& 2671.40&   0.125(11)\\
                     && 2629.92&   0.077(7)\\
                     && 2598.50&   0.040(6)\\
 \hline \end{tabular} \end{center}
 \newpage
{\footnotesize\hspace{35mm}
{\em Table 1 (continued)}\\}
\begin{center}
\begin{tabular}{|l|c|l|l|}  \hline
$E_1$,keV&  $E_i$, keV &$E_2$, keV &
$i_{\gamma\gamma}$\\ \hline
                    && 2568.67&   0.057(11)\\
                     && 2437.54&   0.035(6)\\
                     && 2364.32&   0.060(6)\\
 2904.30& 2679.60(50)& 2679.60&   0.030(8)\\
                     && 2638.12&   0.035(6)\\
                     && 2445.74&   0.022(6)\\
                     && 2280.58&   0.072(6)\\
 2896.80& 2687.10(14)& 2645.62&   0.032(6)\\
                     && 2614.20&   0.026(6)\\
 2893.70& 2690.20(35)& 2648.72&   0.028(6)\\
                     && 2456.34&   0.033(7)\\
 2890.00& 2693.90(52)& 2621.00&   0.026(6)\\
                     && 2591.17&   0.019(6)\\
 2886.90& 2697.00(50)& 2594.27&   0.047(8)\\
                     && 2463.14&   0.030(7)\\
                     && 2401.32&   0.051(8)\\
                     && 2389.92&   0.019(6)\\
 2884.40& 2699.50(70)& 2699.50&   0.076(8)\\
                     && 2658.02&   0.041(7)\\
                     && 2626.60&   0.024(6)\\
                     && 2596.77&   0.101(10)\\
                     && 2465.64&   0.024(7)\\
                     && 2300.48&   0.026(5)\\
 2880.20& 2703.70(75)& 2662.22&   0.046(7)\\
                     && 2600.97&   0.032(8)\\
                     && 2469.84&   0.076(9)\\
                     && 2396.62&   0.045(6)\\
                     && 2304.68&   0.016(5)\\
 2875.00& 2708.90(65)& 2475.04&   0.066(8)\\
                     && 2413.22&   0.022(7)\\
 2869.10& 2714.80(14)& 2480.94&   0.022(6)\\
                     && 2407.72&   0.032(6)\\
 2867.00& 2716.90(42)& 2409.82&   0.029(6)\\
 \hline \end{tabular} \end{center}
 \newpage
{\footnotesize\hspace{35mm}
{\em   }\\}
\begin{center}
\begin{tabular}{|l|c|l|l|}  \hline
$E_1$,keV&  $E_i$, keV &$E_2$, keV &
$i_{\gamma\gamma}$\\ \hline
 2863.70& 2720.20(60)& 2720.20&   0.025(6)\\
                     && 2678.72&   0.057(12)\\
                     && 2647.30&   0.026(6)\\
                     && 2486.34&   0.026(6)\\
                     && 2413.12&   0.028(6)\\
                     && 2321.18&   0.021(5)\\
                     && 2264.43&   0.024(7)\\
 2860.30& 2723.60(36)& 2650.70&   0.023(6)\\
                     && 2489.74&   0.048(6)\\
 2855.70& 2728.20(52)& 2686.72&   0.057(7)\\
                     && 2625.47&   0.029(6)\\
 2851.80& 2732.10(42)& 2732.10&   0.062(11)\\
 2849.60& 2734.30(18)& 2734.30&   0.062(10)\\
                     && 2500.44&   0.027(6)\\
 2845.50& 2738.40(40)& 2696.92&   0.018(10)\\
                     && 2635.67&   0.021(8)\\
                     && 2504.54&   0.021(6)\\
 2842.00& 2741.90(20)& 2741.90&   0.056(6)\\
                     && 2639.17&   0.030(8)\\
                     && 2508.04&   0.020(6)\\
                     && 2434.82&   0.021(6)\\
 2837.20& 2746.70(12)& 2643.97&   0.053(8)\\
 2834.10& 2749.80(22)& 2749.80&   0.025(6)\\
                     && 2676.90&   0.056(6)\\
                     && 2647.07&   0.038(8)\\
 2831.00& 2752.90(36)& 2519.04&   0.016(6)\\
                     && 2445.82&   0.027(6)\\
 2825.70& 2758.20(29)& 2524.34&   0.044(6)\\
 2822.20& 2761.70(85)& 2761.70&   0.030(6)\\
 2819.00& 2764.90(30)& 2457.82&   0.026(6)\\
 2810.00& 2773.90(30)& 2773.90&   0.040(6)\\
                     && 2732.42&   0.018(6)\\
                     && 2671.17&   0.070(8)\\
 \hline \end{tabular} \end{center}
 \newpage
{\footnotesize\hspace{35mm}
{\em Table 1 (continued)}\\}
\begin{center}
\begin{tabular}{|l|c|l|l|}  \hline
$E_1$,keV&  $E_i$, keV &$E_2$, keV &
$i_{\gamma\gamma}$\\ \hline
                     && 2540.04&   0.022(6)\\
                     && 2466.82&   0.033(6)\\
 2804.50& 2779.40(30)& 2545.54&   0.018(6)\\
 2801.80& 2782.10(38)& 2782.10&   0.025(6)\\
                     && 2740.62&   0.023(6)\\
                     && 2709.20&   0.032(6)\\
                     && 2679.37&   0.030(6)\\
                     && 2548.24&   0.029(9)\\
                     && 2475.02&   0.023(6)\\
                     && 2326.33&   0.026(7)\\
 2799.80& 2784.10(44)& 2550.24&   0.027(9)\\
                     && 2477.02&   0.041(6)\\
 2791.90& 2792.00(23)& 2750.52&   0.015(6)\\
                     && 2496.32&   0.024(7)\\
                     && 2484.92&   0.049(6)\\
 2786.00& 2797.90(45)& 2797.90&   0.037(6)\\
                     && 2695.17&   0.087(8)\\
                     && 2564.04&   0.022(6)\\
                     && 2490.82&   0.039(6)\\
 2778.40& 2805.50(85)& 2805.50&   0.027(6)\\
                     && 2764.02&   0.082(7)\\
                     && 2702.77&   0.072(6)\\
                     && 2349.73&   0.032(6)\\
 2772.30& 2811.60(60)& 2738.70&   0.024(6)\\
                     && 2504.52&   0.029(6)\\
 2761.10& 2822.80(19)& 2113.60&   0.030(7)\\
 2753.60& 2830.30(45)& 2727.57&   0.036(6)\\
                     && 2596.44&   0.097(7)\\
 2749.60& 2834.30(60)& 2731.57&   0.042(6)\\
 2727.60& 2856.30(50)& 2856.30&   0.138(9)\\
                     && 2753.57&   0.036(6)\\
                     && 2549.22&   0.032(6)\\
 2720.10& 2863.80(17)& 2822.32&   0.041(6)\\
 \hline \end{tabular} \end{center}
 \newpage
{\footnotesize\hspace{35mm}
{\em       }\\}
\begin{center}
\begin{tabular}{|l|c|l|l|}  \hline
$E_1$,keV&  $E_i$, keV &$E_2$, keV &
$i_{\gamma\gamma}$\\ \hline
                    && 2761.07&   0.055(6)\\
                     && 2408.03&   0.082(7)\\
                     && 2154.60&   0.045(7)\\
 2713.90& 2870.00(22)& 2870.00&   0.071(8)\\
                     && 2828.52&   0.099(9)\\
                     && 2767.27&   0.190(10)\\
 2708.10& 2875.80(40)& 2875.80&   0.045(6)\\
                     && 2773.07&   0.019(6)\\
                     && 2641.94&   0.020(6)\\
 2703.90& 2880.00(44)& 2807.10&   0.027(6)\\
                     && 2646.14&   0.025(6)\\
 2696.90& 2887.00(61)& 2814.10&   0.127(9)\\
                     && 2579.92&   0.019(6)\\
                     && 2487.98&   0.028(6)\\
 2679.80& 2904.10(47)& 2597.02&   0.034(6)\\
 2674.90& 2909.00(92)& 2909.00&   0.026(8)\\
                     && 2867.52&   0.153(12)\\
                     && 2806.27&   0.116(8)\\
                     && 2675.14&   0.069(6)\\
 \hline \end{tabular} \end{center}
 \newpage
{\footnotesize\hspace{35mm}
{\em Table 1 (continued)}\\}
\begin{center}
\begin{tabular}{|l|c|l|l|}  \hline
$E_1$,keV&  $E_i$, keV &$E_2$, keV &
$i_{\gamma\gamma}$\\ \hline
                     && 2601.92&   0.030(7)\\
                     && 2509.98&   0.028(5)\\
 2670.60& 2913.30(40)& 2871.82&   0.037(7)\\
                     && 2606.22&   0.043(7)\\
 2665.90& 2918.00(31)& 2918.00&   0.021(6)\\
                     && 2610.92&   0.021(6)\\
 2611.50& 2972.40(28)& 2930.92&   0.121(10)\\
                     && 2516.63&   0.047(6)\\
 2604.00& 2979.90(29)& 2938.42&   0.060(7)\\
                     && 2877.17&   0.034(8)\\
 2597.00& 2986.90(90)& 2914.00&   0.025(6)\\
 2582.20& 3001.70(30)& 3001.70&   0.024(6)\\
                     && 2898.97&   0.025(6)\\
 2577.30& 3006.60(15)& 3006.60&   0.070(8)\\
                     && 2965.12&   0.026(7)\\
                     && 2903.87&   0.076(8)\\
                     && 2772.74&   0.016(6)\\
 2573.50& 3010.40(100)& 3010.40&   0.035(8)\\
                     && 2937.50&   0.042(5)\\
 \hline \end{tabular} \end{center}
\hspace*{-85mm}{
\parbox{170mm}{
 1.~The lower estimation of $i_{\gamma\gamma}$ for cascades with $E_1<520$
keV or $E_2<520$ keV.

2.~Only statistical uncertainty of determination of energy and intensity.
}}

\onecolumn
\begin{center}
{\sl Table~2. The sum energies (keV) of cascades $E_c$, the calculated
$I^{cal}_{\gamma\gamma}$ and experimental $I^{exp}_{\gamma\gamma}$
intensities (\% per decay) of the two-step cascades in $^{193}Os$.}
\end{center}
\begin{center}
\begin {tabular}{|l|l|l|l|l|l|} \hline
 $E_c$ &
$I^{exp}_{\gamma\gamma}$ &
$I^{cal}_{\gamma\gamma}$ [18,20]&
$I^{cal}_{\gamma\gamma}$ [19,20]&
$I^{cal}_{\gamma\gamma}$ [18,21]&
$I^{cal}_{\gamma\gamma}$ [19,21]\\ \hline
 5583.90& 12.5(2) & 6.2 & 6.5&4.7 &5.3\\
 5542.42& 12.2(2) &5.6 & 5.7&4.4 & 4.7\\
 5511.00& 4.5(1) & 2.9 & 3.0& 2.2 & 2.5 \\
 5481.17& 19.0(1) & 5.1 & 5.3 &4.0 & 4.4\\
 5350.04& 7.2 (2) & 4.1 & 4.4  & 3.2 & 3.7\\
 5288.22& 3.9(2) & 2.0 & 2.2 & 1.6 & 1.8\\
 5276.82& 8.1(4) & 3.6 & 4.0 & 2.9 & 3.4\\
 5184.88& 4.0(2) & 1.6 & 1.9 & 1.3 & 1.6\\
 5128.13& 3.8(2) & 1.5 & 1.8 & 1.2 & 1.5\\
 4874.70& 1.5(1) & 0.9 & 1.2 & 0.7 & 1.0 \\
 4694.42& 3.4(2) & 1.1 & 1.7& 1.0 & 1.5 \\
total& 81(1) & 34.6 & 37.7 & 27.2 & 31.4\\
\hline
\end{tabular}
\end{center}

\begin{center}
{\sl Table~3. The energy (keV) $E_c$ of the experimentally unresolved
cascades, the energy (keV) $E_f$ of the corresponding them final levels
with the probable values of $J^{\pi}$ for $^{191,193}Os$.}
\end{center}
\begin{center}
\begin {tabular}{|l|l|l|l|l|l|} \hline
\multicolumn{3}{|c|} {$^{191}Os$} &
\multicolumn{3}{|c|} {$^{193}Os$} \\ \hline
        $E_c$ & $E_f$ & $J^{\pi}$ & $E_c$ & $E_f$ & $J^{\pi}$\\ \hline
         5485.9 &273&  5/2- &             5481.2   &103&   3/2-\\
         5287.0   &472&  5/2- &           5288.2   &296&   5/2-\\
         5184.5 &574&  1/2-  3/2-   &     5184.9   &399&   5/2-\\
         5127.9 &631& 5/2-        &       5128.1  &456&    5/2-\\
\hline
\end{tabular}
\end{center}

\newpage
\begin{flushleft}
{\large\bf References}
\end{flushleft}
\begin{flushleft}
\begin{tabular}{@{}r@{ }p{5.65in}}
$[1]$ & S. T. Boneva, E. V. Vasilieva, Yu. P. Popov, A. M. Sukhovoj
 and V. A. Khitrov, Sov. J. Part. Nucl. {\bf 22(2)} (1991) 232;\\
$[2]$ & S.T. Boneva et al., Part. Nucl. {\bf 22(6)}, 698 (1991) 698;\\
$[3]$ & E. P. Grigoriev, V. A. Khitrov, A. M. Sukhovoj and E. V. Vasilieva
Fizika B (Zagreb)  {\bf 9(4)} (2000) 147;\\
$[4]$ & J. Honz\'atko  et al., Nucl. Instr. and Meth. {\bf A376}, 434 (1996)\\
$[5]$ & F.D. Corte, A. Simonits, in {\it Proceedings of International Conference
on Nuclear Data for Science and Technology, Mito, 1988}, edited by S. Igarasi
(Japan Atomic Energy Research Institute, 1988), 583;\\
$[6]$ & A.M. Sukhovoj, V.A. Khitrov, Sov. J. Prib. Tekhn. Eksp. {\bf 5},
(1984) 27;\\
$[7]$ & Yu.P. Popov, A. M. Sukhovoj, V. A. Khitrov and Yu. S. Yazvitsky,
 Izv. AN SSSR, Ser. Fiz. {\bf 48}, (1984) 891;\\
$[8]$ & S.T.Boneva, E.N.Vasilieva and  A.M.Sukhovoj, Izv. RAN., ser. fiz.,
{\bf 51(11)}  (1989) 2023\\
$[9]$ & E. Browne, Nuclear Data Sheets, 1989, {\bf V.56}, (1989) 709;\\
$[10]$ & P. Fettweis and J.C. Dehaes, Z. Phys. {\bf 314}, (1983) 159;\\
$[11]$ & R.F. Casten  et al., Nucl. Phys. {\bf A316}, (1979) 61;\\
$[12]$ & Mughabghab S.F., {\it Neutron Cross Sections. Part B.}, NY: Academic
Press 1984\\
$[13]$ & V.A. Bondarenko, V.A.Khitrov, A.M.Sukhovoj, J.Honzatko and I.Tomandl,
 JINR preprint E3-99-343, Dubna, 1999; \\
$[14]$ &  D.Benson, Jr., P.Kleinheinz, R.K.Sheline and E.B.Shera
Z.Phys. A285,  (1978) 405\\
$[15]$ & http://www.nndc.bnl.gov/wallet/tnc/capgam.shtml\\
$[16]$ & A.M. Sukhovoj and V.A. Khitrov, Physics of Atomic Nuclei {\bf 62(1)},
(1999) 19;\\
$[17]$  & C.F. Porter and R.G. Thomas, Phys. Rev. 1956. {\bf
104}, (1956) 483;\\
$[18]$ & W. Dilg, W. Schantl, H. Vonach and M. Uhl, Nucl. Phys {\bf A217} (1973)
269;\\
$[19]$ & A.V. Ignatyuk, Proc. of IAEA Consultants Meeting on the use of Nuclear
Theory and Neutron Nuclear Data Evaluation (Trieste, 1975)
IAEA-190, Vol. 1, (1976) 211; \\
$[20]$ & P. Axel,
Phys. Rev. {\bf 126}, (1962) 683;\\
$[21]$ & S.G. Kadmenskij, V.P.
Markushev and W.I. Furman, Yad. Fiz.  {\bf 37}, (1983) 165;\\
$[22]$ & E. V. Vasilieva,
Sukhovoj A.M. and Khitrov V.A.,
 Phys. of Atomic Nuclei, {\bf 64(2)} (2001) 153;\\
$[23]$ & L.A.Malov and V.G.Soloviev, Yad. Fiz., {\bf 26(4)} (1977) 729;\\
$[24]$ & A.M. Sukhovoj and V.A. Khitrov, Izv. RAN, Ser. Fiz. {\bf 61(11)} (1997) 2068;\\
$[25]$ & E. V. Vasilieva et al., Bulletin of the Russian Academy of Science,
Physics {\bf 57} (1993) 1582;\\
$[26]$ & S.T. Boneva et al., Phys. of Atomic Nuclei, {\bf 62(5)} (1999) 832;\\
\end{tabular}
\end{flushleft}
\newpage
{\bf Figure captions}\\
{\sl Fig.~1.~The part of the sum coincidence spectrum for $^{191,193}Os$.
The peaks are  labelled with the energy (in keV) of final cascade levels.
The mass of the corresponding isotope is given in brackets.}

{\sl Fig.~2.~The part of the intensity distribution of the two-step cascades
summed over 11 cascade final levels in $^{193}Os$.}

{\sl Fig.~3.~Frequency distribution of differences of the primary transition
energies  $E_1$ of the cascades in $^{193}Os$ and $^{191}Os$ [13].
Horizontal lines represent the average value and mean-square deviation from
it.}

{\sl Fig.~4.~The cumulative cascade intensities in $^{193}Os$ for four
excitation energy intervals 1.50-1.75, 2.00-2.25, 2.50-2.75, and 2.75-3.00 MeV
versus intensity (histograms). Approximation and extrapolation of cumulative
intensities to values corresponding to $I_{\gamma\gamma}=0$ are illustrated
by solid lines.}

{\sl Fig.~5.~The number of the observed intermediate levels of cascades in
$^{193}Os$ (Table 1) for the excitation energy interval of 100 keV (circles).
Curves 1 and 2 represent the predictions of the models [18] and [19],
respectively. The histogram is the estimation [16] of the level density from
the shape of the distribution of cumulative sums of cascade intensities.}

{\sl Fig.~6.~The dependence of the ``smoothed" intensities  of
resolved  cascades listed in Table 1 on the excitation energy. Possible
``bands" of practically harmonic excitations of the nucleus are marked.
The parameter $\sigma =25$ keV was used.}

{\sl Fig.~7.~The values of the functional $A(T)$ for three registration
thresholds of most intense cascades. The value of the registration
threshold (\% per decay) is given in the figure.}

{\sl Fig.~8.~The value of the equidistant period $T$ for $^{193}Os$ (asterisk),
even-odd (triangles) and odd-odd (circles) nuclei as a function of the
number of boson pairs, $N_b$ in unfilled shells. The line represents
possible dependence (drawn by eye).}
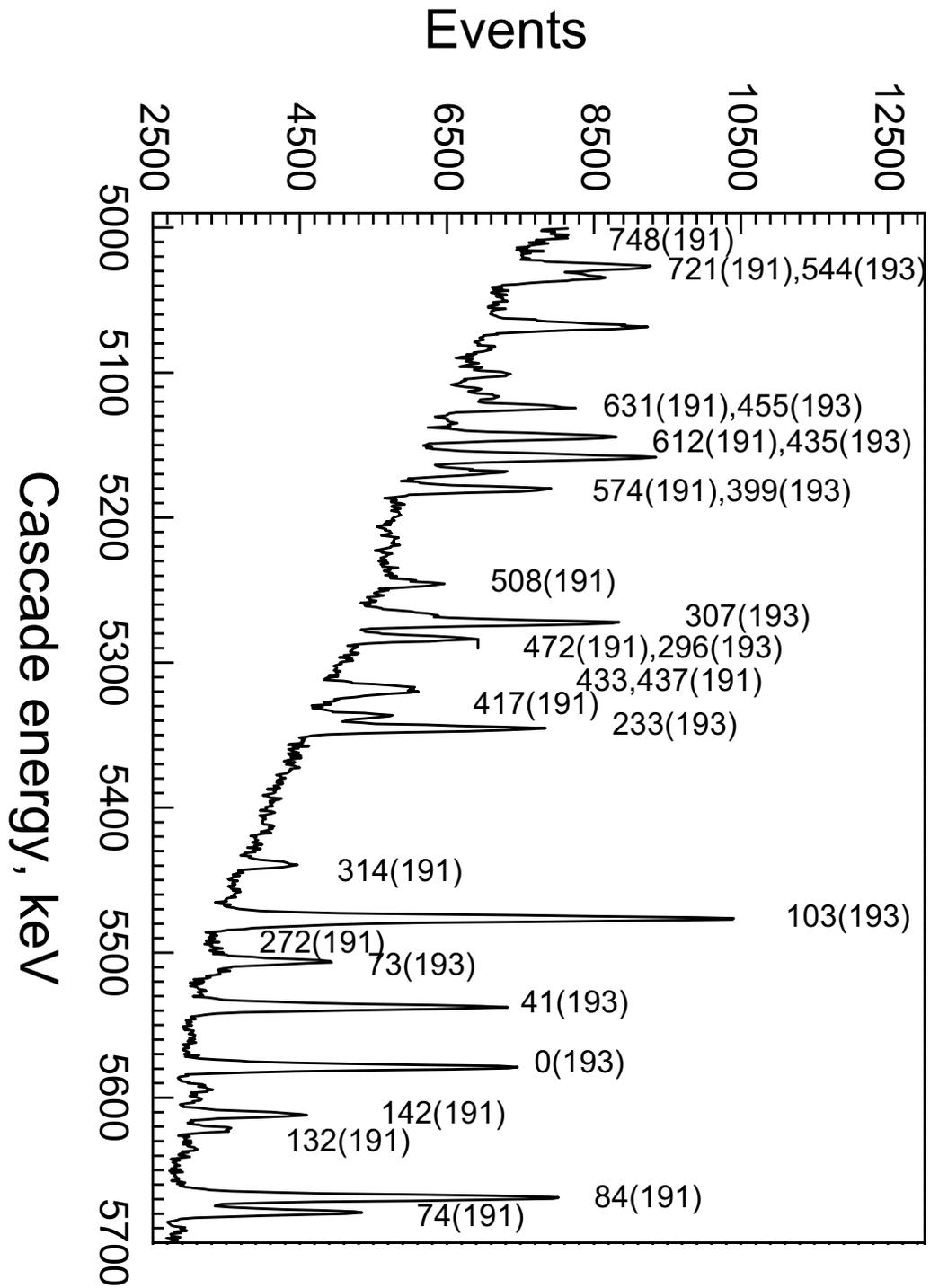
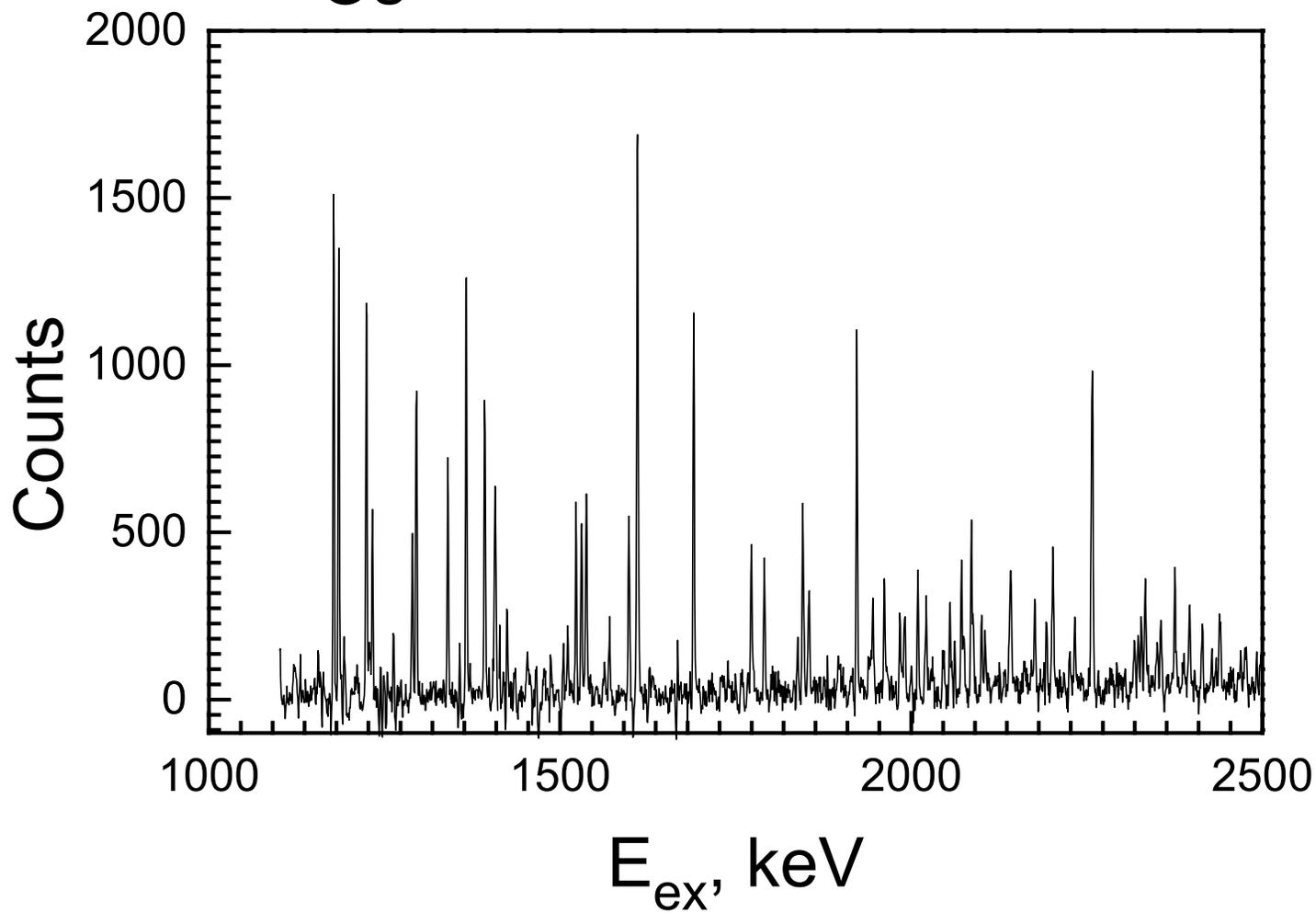
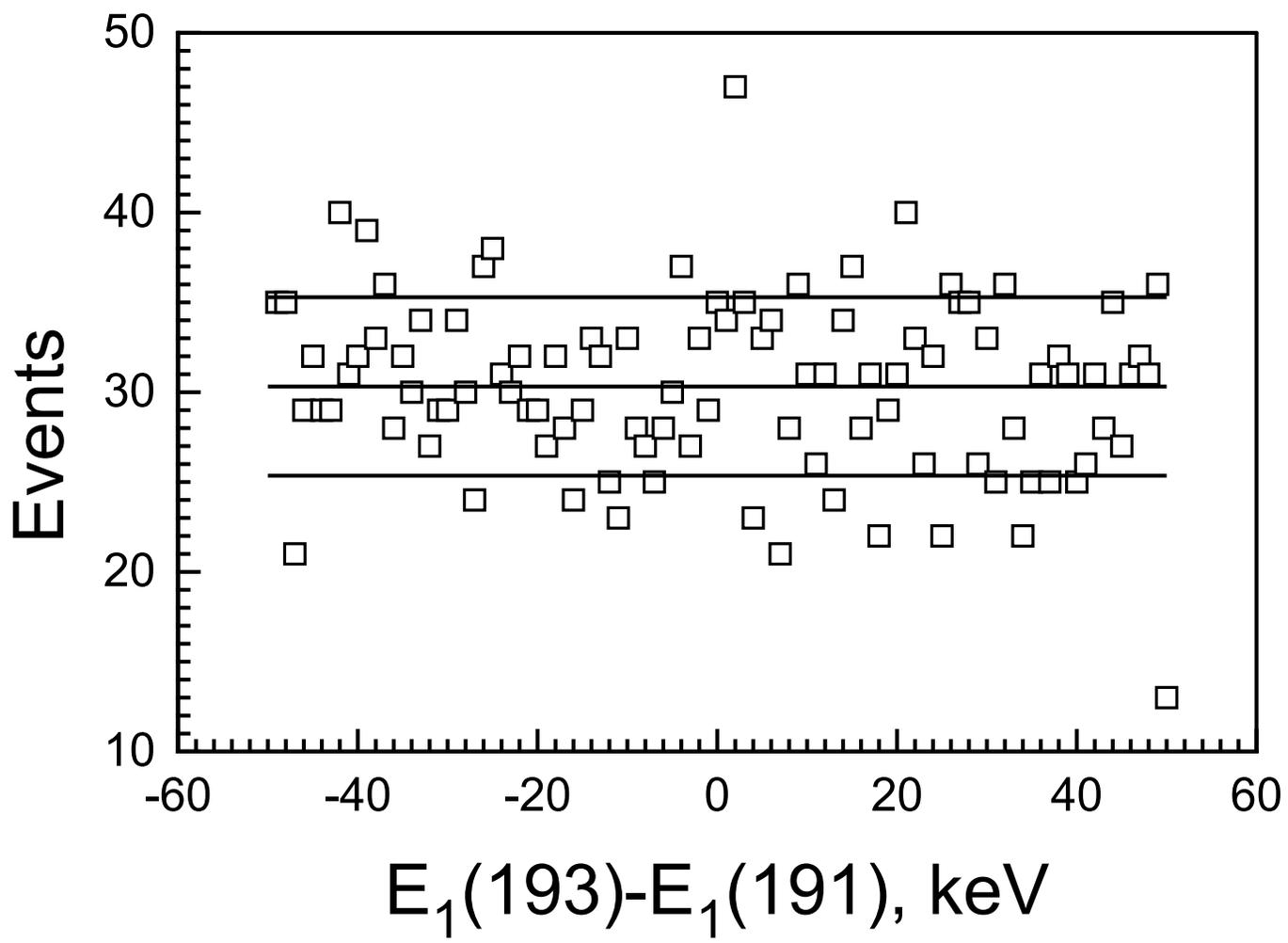
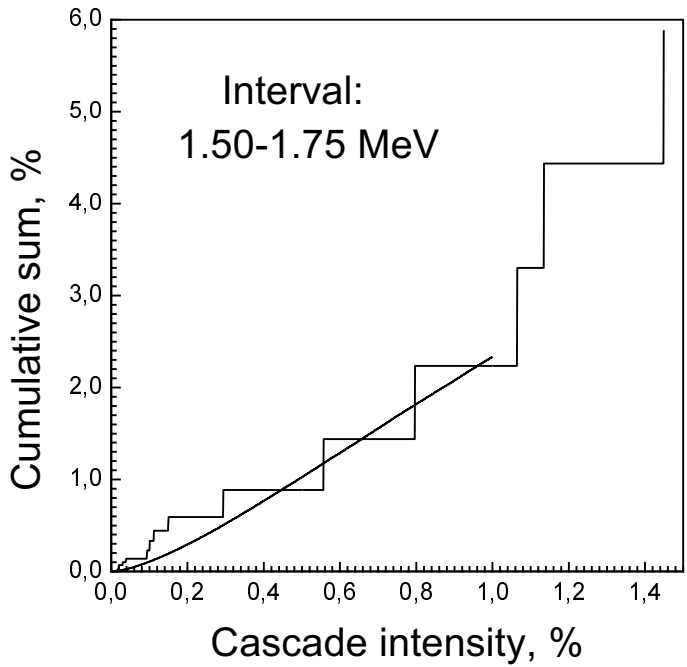
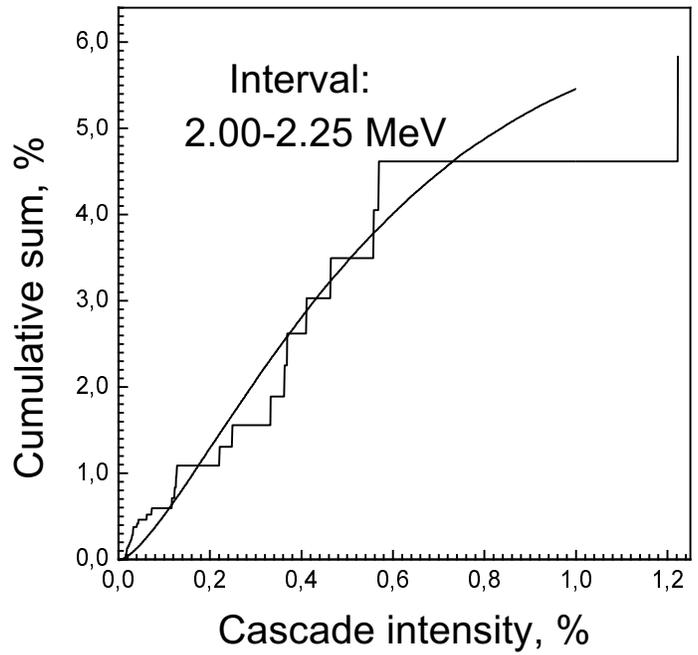
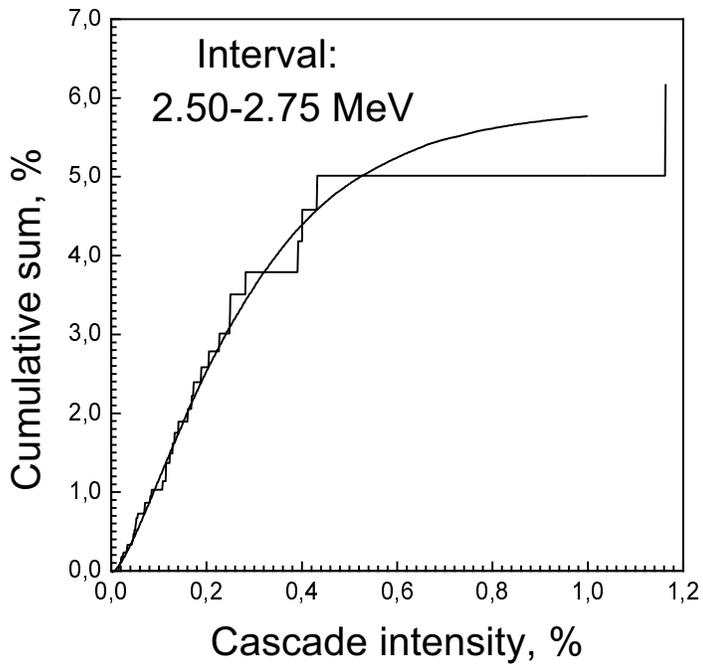
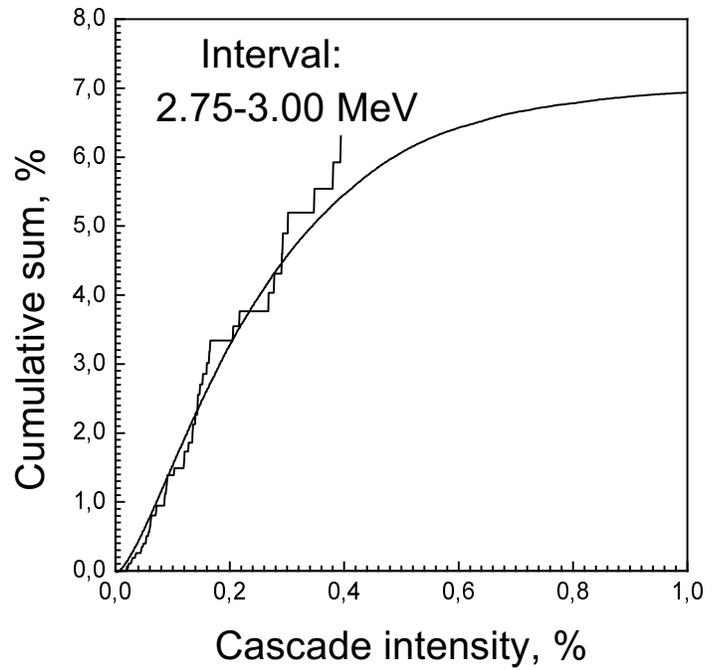
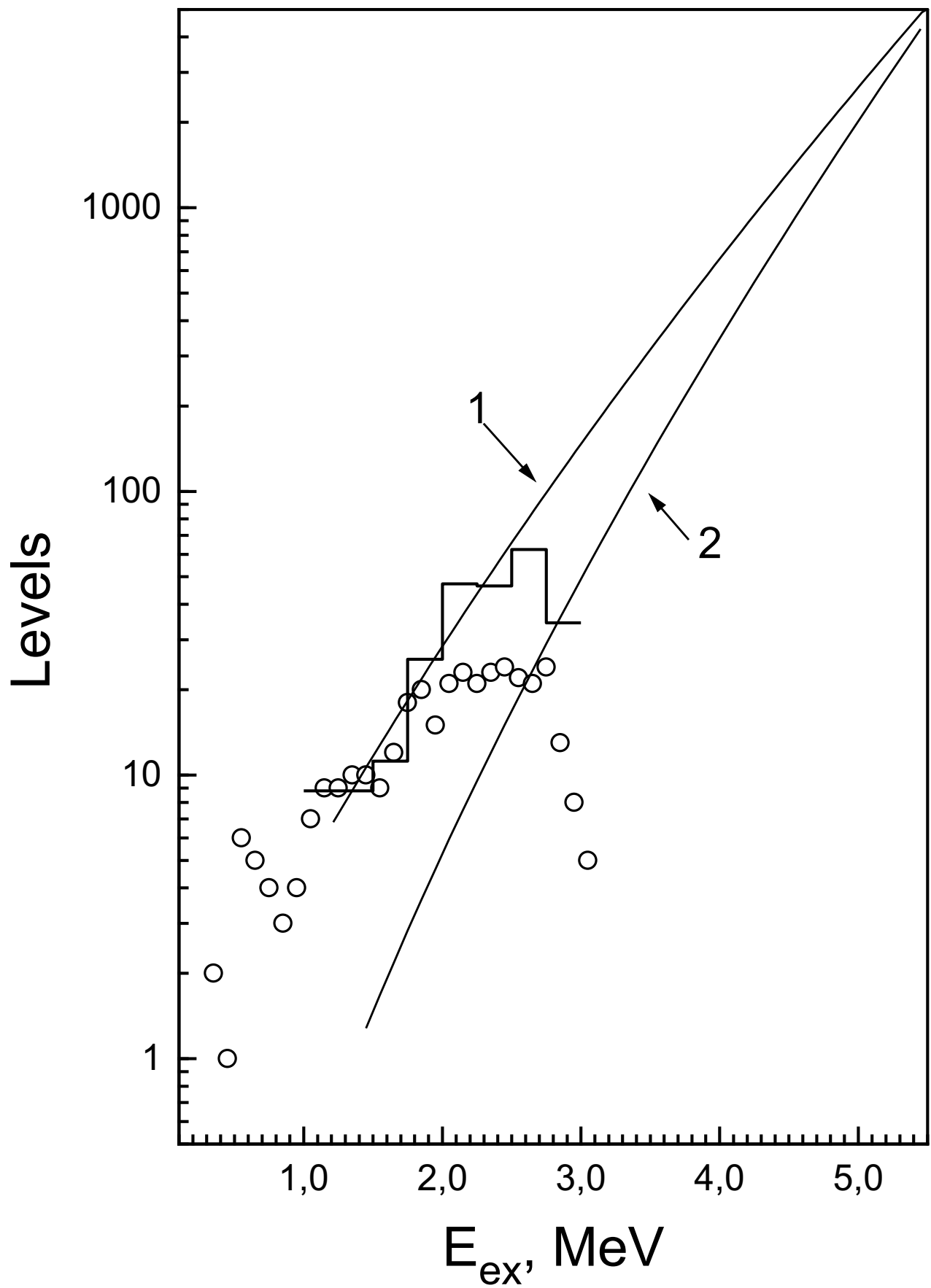
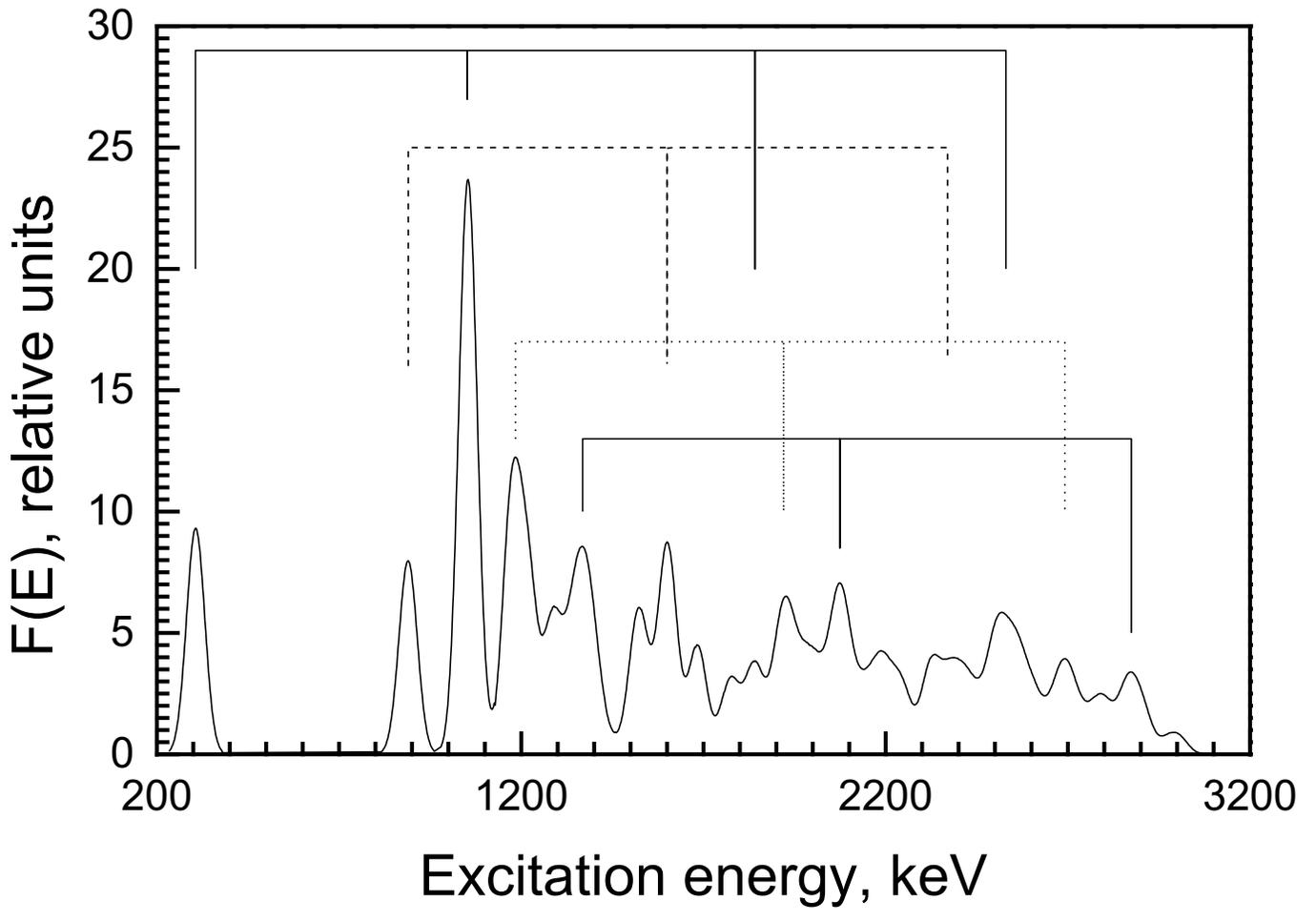
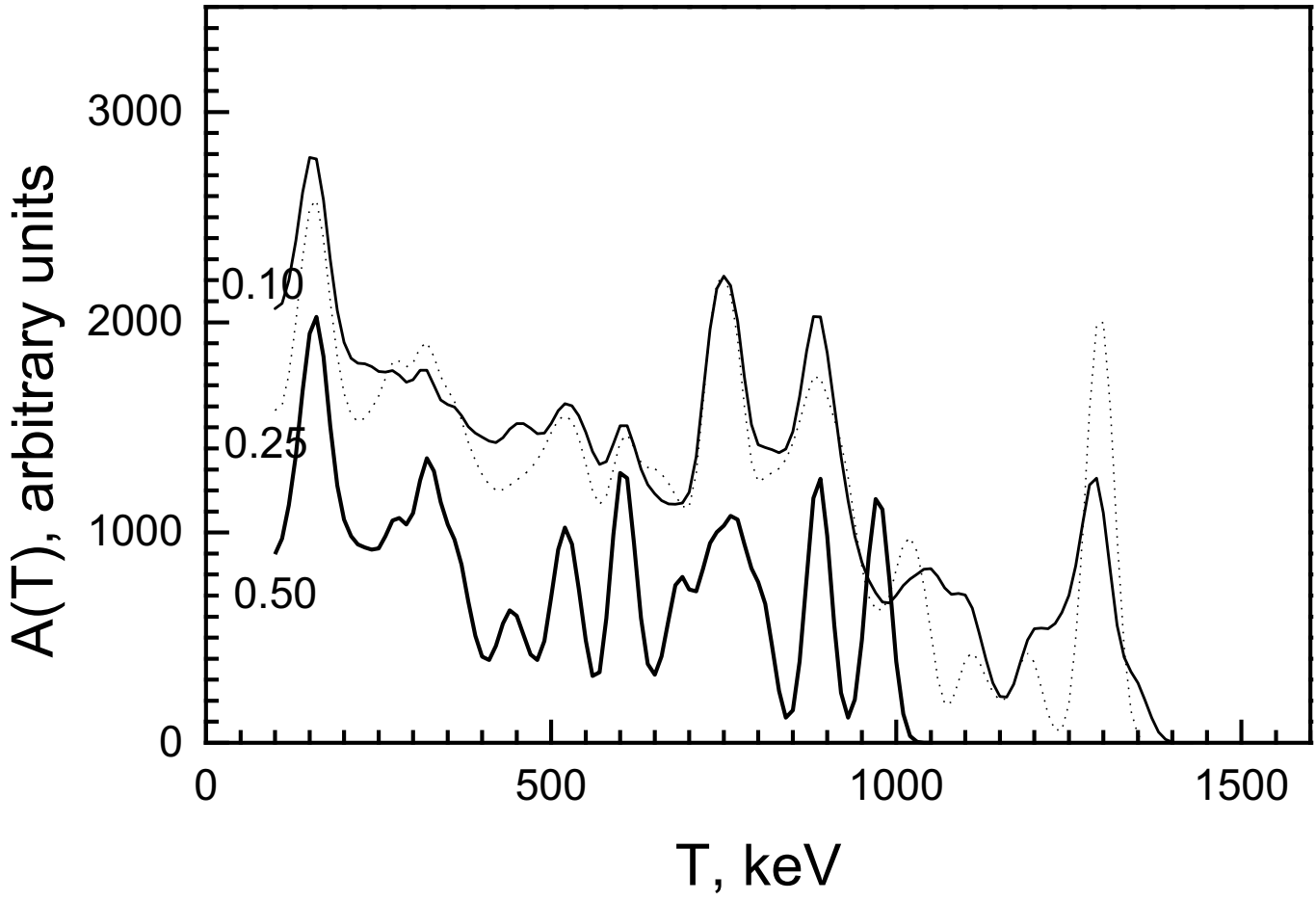
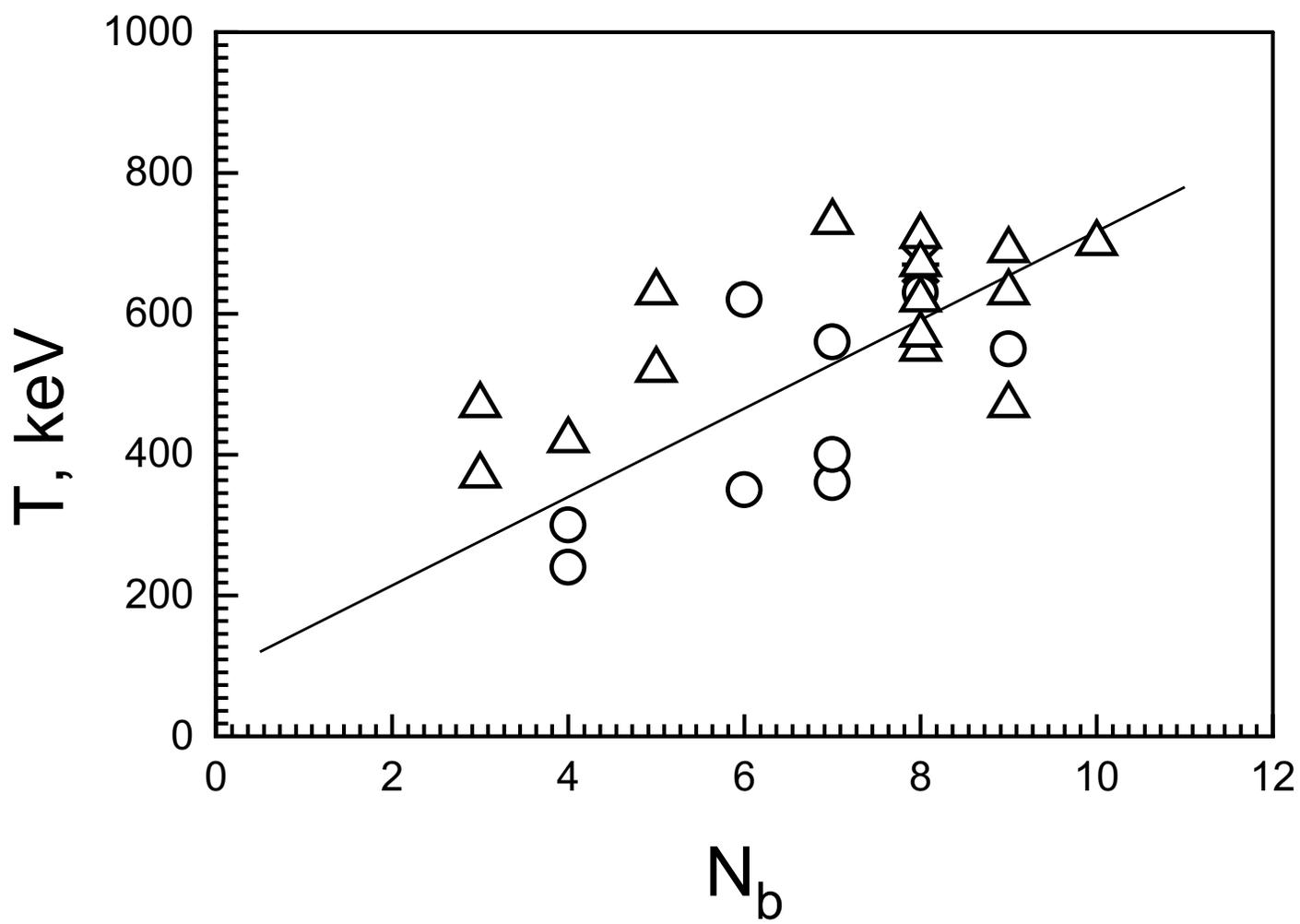
\end{document}